\documentclass[prl,twocolumn,superscriptaddress,showpacs,floatfix,amsmath,amssymb]{revtex4-2}
\usepackage{dcolumn}
\usepackage{xcolor}
\usepackage{braket}
\usepackage[utf8]{inputenc}
\usepackage{amsmath}
\usepackage{amssymb}
\usepackage{dsfont}
\usepackage{physics}
\usepackage{color}
\usepackage{graphicx}
\usepackage{mathtools}
\usepackage{overpic}
\usepackage[toc,page]{appendix}
\usepackage{chemfig}
\usepackage{ccycle}
\usepackage{comment}
\usepackage{textcomp}
\usepackage{gensymb}
\usepackage{graphics}
\usepackage{epsfig}
\usepackage{subfigure}
\usepackage{siunitx}
\usepackage[T1]{fontenc}
\usepackage{soul}
\usepackage{mathrsfs}

\usepackage{hyperref}
\hypersetup{
    colorlinks=true,
    linkcolor=cyan,
    urlcolor=cyan,
    citecolor=cyan,
    }

\begin{document}
\preprint{APS}

\title{Disorder-Driven Enhancement of Coulomb Repulsion Governs The Superconducting Dome in Ionic-Liquid-Gated Quasi-2D Materials}

\author{Giovanni Marini}
\affiliation{Dipartimento di Fisica, Universit\'a di Trento, via Sommarive 14, I-38123 Povo, Italy}
\affiliation{Graphene Labs, Fondazione Istituto Italiano di Tecnologia, Via Morego, I-16163 Genova, Italy}

\author{Pierluigi Cudazzo}
\affiliation{Dipartimento di Fisica, Universit\'a di Trento, via Sommarive 14, I-38123 Povo, Italy}

\author{Matteo Calandra}
\affiliation{Dipartimento di Fisica, Universit\'a di Trento, via Sommarive 14, I-38123 Povo, Italy}
\affiliation{Graphene Labs, Fondazione Istituto Italiano di Tecnologia, Via Morego, I-16163 Genova, Italy}

\begin{abstract}
The superconducting dome in the $T_{c}$ versus doping phase diagram, found in cuprates, nickelates, twisted bilayer graphene, and transition metal dichalcogenides, is often considered a signature of unconventional pairing. Identifying the underlying mechanisms of any of these phase diagrams and developing a reliable theoretical understanding of it remains a critical challenge. Here we demonstrate that, in ionic-liquid-gated quasi-2D materials, the disordered ionic potential from the frozen ionic liquid drives the system close to Anderson transition. In this regime, quenched charge fluctuations and reduced screening markedly enhance repulsive Coulomb interactions, suppressing $T_{c}$ and naturally leading to the formation of a superconducting dome. By integrating a many-body approach including disorder with first-principles calculations, we obtain the phase diagrams and tunneling spectra of gated few-layers transition metal dichalchogenides in robust quantitative agreement with experiments. Our findings establish that disorder-driven enhancement of Coulomb repulsion is a fundamental feature of ionic-liquid-gated quasi-2D materials at high bias.
\end{abstract}

\maketitle

% The first paragraph of any Science paper does NOT have a heading
% Nor is it indented

Ionic-liquid based electric double layer field effect transistors constitute a unique platform to achieve high doping in low dimensional semiconductors and insulators: the charged ions in the electrolyte form an electric double-layer at the interface with the semiconductor, inducing charge accumulation in the latter \cite{https://doi.org/10.1002/adfm.200801633}. A gate-induced superconductivity dome has been shown to occur  cuprates \cite{Bollinger2011,Sterpetti2017}, twisted bilayer graphene \cite{Cao2018} and gated few-layer transition metal dichalcogenides (TMD) where superconductivity was first reported in molybdenum disulfide (MoS$_2$) by Ye {\it et al.} in 2012\cite{doi:10.1126/science.1228006} and soon after in by other groups and in other TMDs\cite{PMID:26235962,doi:10.1073/pnas.1716781115,Costanzo2018,AliElYumin2019,doi:10.1021/acs.nanolett.8b01390}.

In all these materials, the superconducting dome feature is unexplained, qualitatively and quantitatively. In most of the cases, this is due to the difficulties in determining the pairing mechanism. Gated (electron-doped) few layers TMDs are an exception as there is evidence that superconductivity is mediated by phonons\cite{PhysRevB.87.241408,PhysRevB.90.245105,Marini_2023}, however the most accurate calculations predict a linear growing of $T_{c}$ with voltage and neither a saturation nor a superconducting dome \cite{Marini_2023}.
Some theoretical works attributed the saturation of $T_{c}$ at high voltage to a charge density wave (CDW) \cite{PhysRevB.87.241408,PhysRevB.90.245105, GirottoErhardt2025}. However, this has been recently disproved theoretically\cite{Marini_2023} and 
 such a transition has never been detected in gated experiments, while in theoretical calculations the CDW arises artificially from three approximations: (i) the inappropriate modeling of field-effect doping by employing a uniform charged background, (ii) the neglect of anharmonic effects, (iii) the assumption that the sample thickness is irrelevant. These approximations result in excessively low phonon frequencies, unrealistically high electron-phonon coupling, wrong thickness dependence\cite{Fu2017}, incorrect electronic band structure. 
For example, in Ref. \cite{GirottoErhardt2025} these approximations has lead to a factor of three overestimation in $T_{c}$ at an electron doping $n_e=2.2\times 10^{-14}~e^-$/cm$^{2}$ and a $T_{c}$ versus doping diagram in quantitative disagreement with experiments (see also Fig. S8 in the Supplemental Material (SM)), with superconductivity occurring no earlier than at $n_e=1.5\times  10^{-14}e^-$/cm$^{2}$ and only in proximity of the artificial CDW due to the aforementioned modeling problems; conversely, in experiments it occurs at $n_e\approx 0.5\times  10^{-14} e^-$/cm$^{2}$. %\st{Finally, in experiments the $T_{c}$ versus doping curve is smooth, while in these calculations it has a discontinuous shape at the artificial CDW transition.}

All previous theoretical calculations neglected an essential point, namely the crucial role of disorder. When the flake is gated at temperatures where the electrolyte (typically, DEME-TFSI in experiments on TMDs) remains in the liquid state, an electric double layer forms, characterized by the accumulation of positively charged molecules ( DEME$^+$, N,N-diethyl-N-methyl-N-(2-methoxyethyl)ammonium) at the surface of the gated material. As the temperature is lowered, the electrolyte freezes in a disordered ionic configuration\cite{SATO20043603,MCCANN2015252}, and charged impurities occur on the surface of the sample in contact with the electrolyte \cite{doi:10.1073/pnas.1716781115}. The higher the voltage, the larger the impurity concentration.  There are four key experimental findings confirming that the disorder induced by the electrolyte has substantial effects on the superconducting properties of gated TMDs. First, tunneling spectra show an extreme variability of the superconducting gap size if measurements are taken in different parts of the sample (see inset of Fig. 3 (c) in Ref. \cite{Costanzo2018}). Second, tunneling spectra shows a yet unexplained V-shape feature with kinks \cite{Costanzo2018} which cannot be reconciled with  a phonon-mediated superconductor with s-wave pairing in the clean limit \cite{Marini_2023}.
Third, in a single layer WS$_2$ a full superconducting dome as a function of doping is measured, flanked by insulating states on both the underdoped and overdoped sides \cite{doi:10.1073/pnas.1716781115}. The  high-doping insulating state does not occur at any fractional filling and it is, thus, unrelated to any band-insulating or Mott-insulating state; rather, it is strongly reminiscent of an Anderson localization. Four, the region of the dome where $T_{c}$ decreases corresponds to a strong reduction of the measured low-temperature Hall mobility \cite{doi:10.1073/pnas.1716781115,Saito2016,doi:10.1126/science.1228006}, suggestive of a competition between superconductivity and disorder. These experimental findings point towards a superconducting state where strong disorder plays an increasing important role as the voltage is increased.

The effect of disorder in low dimensional superconductors mainly follows in two regimes. If the elastic scattering introduced by disorder is larger than the superconducting gap, but the impurity concentration is not high enough to push the system close to an Anderson localization, then $T_{c}$ is weakly affected by disorder. This regime corresponds to the limit in which $k_F l \gg 1$, $k_F$ being the Fermi momentum and $l$ the mean free path. Under these conditions, within the BCS theory\cite{PhysRev.108.1175}, the pairing does not occur between states having opposite electron-momenta, but between a disordered state and its time reversed state \cite{ANDERSON195926}. This regime has been thoroughly investigated by Gor'kov in Ref. \cite{Gorkov}, where it was demonstrated that the density of states is affected at order $( k_F l )^{-1}$. 
Charge fluctuations are not quenched and screening is practically unaffected. The Coulomb vertex is almost unchanged and the repulsion between electrons is essentially the same as in the clean limit.

%The Cooper pairs attraction generated by the electron-phonon coupling is ruled by two quantities, namely the disorder induced smearing of the Fermi surface (order of  $( k_F l )^-1$) and the ratio between the phonon wavelength and the mean-free path. Tipically, random diluted scatterers do not influence long wavelength phonon states and, consequently, $T_{c}$ weakly differs from the clean limit. This regime can be in principle addressed by employing large enough supercells in first principles calculations to account for the diluteness of the impurities.

The situation changes qualitatively when the disorder becomes sufficiently strong that $ (k_F l)^{-1} \approx 1$ and the system approaches Anderson localization, while still remaining non-insulating. In the quasi-two-dimensional case represented by few-layer-thick samples, this regime can be approached easily as  localization can occur at relatively low values of impurity concentration\cite{PhysRevLett.42.673}.  The charge fluctuations are strongly quenched and the kinetic energy is strongly reduced. As a consequence,  Coulomb repulsion and screening are strongly influenced and quantum fluctuations become crucial\cite{PhysRevB.28.117,doi:10.1143/JPSJ.51.1380,FINKELSTEIN1994636,Eckern1988}.  In this scenario, corrections to the superconducting gap equation of the order of the the product $(\varepsilon_F \tau_0)^{-1}$ become important, $\varepsilon_F$ being the Fermi energy and $\tau_0$ the average scattering time. This effect determines the suppression of superconducting critical temperature for thin metallic films\cite{PhysRevB.1.1078} in combination with the observation of exponentially increasing resistivity.
Theoretical analysis pointing out the combined role of disorder and dimensionality has been given in many theoretical works for thick flakes \cite{1973JETP} and in the two dimensional limit\cite{doi:10.1143/JPSJ.51.1380,TAKAGI1982643,1987PZETF4537F,FINKELSTEIN1994636}. These remarkable theoretical developments, rest mainly on the BCS model in the presence of disorder and Coulomb repulsion. They have never been coupled to the high accuracy of first principles calculations to describe the superconducting phase in the presence
of disorder in quasi-two-dimensional systems.

In this work we substantiate the idea that disorder plays a critical role for $T_{c}$ suppression in gated systems at high voltage. By extending previous many-body approaches \cite{doi:10.1143/JPSJ.51.1380, TAKAGI1982643, 1987PZETF4537F,FINKELSTEIN1994636}, we develop a first principles framework to evaluate superconducting properties in the presence of strong disorder occurring in proximity of an Anderson transition. We demonstrate that this approach quantitatively describes the superconducting dome in single-side gated few layer MoS$_2$ and the smaller $T_{c}$ saturation at high voltage in few layers MoSe$_2$. Finally, we explain both the V-shaped form, the anomalous kinks and temperature dependence of the tunneling spectra reaching an unprecedented agreement with experimental data. The present results establish that disorder-driven enhancement of Coulomb repulsion is a universal feature of ionic liquid-gated quasi-2D materials at high bias.

The effect of strong disorder on $T_{c}$ is described by including the perturbative corrections to the pair propagator following the approach presented in Ref.\cite{doi:10.1143/JPSJ.51.1380} for 2D systems and discussed in \cite{FINKELSTEIN1994636,1987PZETF4537F,Larkin2008} (see also the SM, Sec. S3). By considering the multi-valley nature of our system with almost perfect parabolic valleys, we find the following  self-consistent equation for $T_{c}$ \cite{doi:10.1143/JPSJ.51.1380}:
\begin{widetext}
\begin{equation}
    \ln(T_c/T_{c0}) =[ -\dfrac{(g_1 N(0)-3\lambda)}{4\pi\varepsilon_F \tau_0}(\ln \dfrac{1}{T_c \tau_0})^2-\dfrac{(g_1 N(0)+\lambda)}{6\pi\varepsilon_F\tau_0}(\ln\dfrac{1}{T_c\tau_0})^3] \label{eq:Tcisom}
\end{equation}
\end{widetext}

where $N(0$) is the density of states  per spin, $\tau_0$ is the average scattering time, $\varepsilon_F$ is the Fermi energy in the presence of disorder,$g_1$ is a constant that accounts for electron-electron interaction ($g_1 N(0) \approx \frac{g_v}{2}$, where $g_v$ is the valley degeneracy) \cite{FINKELSTEIN1994636}) and $\lambda$ is the electron-phonon interaction (see the SM, Section S3 for a precise description), while $T_{c0}$ is the superconducting critical temperature in the absence of disorder. As the scattering time $\tau_0$ decreases, the right-hand side of Eq.\ref{eq:Tcisom} becomes larger, causing a suppression of $T_{c}$. More detailed insights on the behavior of the superconducting critical temperature $T_{c}$ emerging from the self-consistent solution of Eq.\ref{eq:Tcisom} as a function of the Fermi energy $\varepsilon_F$ and $\tau_0$ are given in the SM, Section S5.

\begin{figure}[t!]
\centerline{\includegraphics[width=0.9\columnwidth]{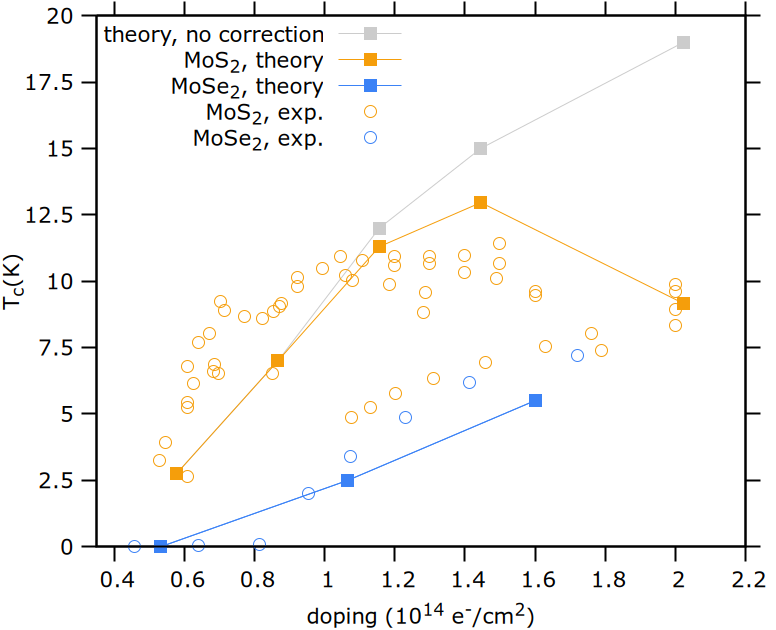}}
\caption{ Predicted superconducting critical temperature $T_{c}$ as a function of doping with disorder+interaction corrections for single-side gated MoS$_2$ (orange filled squares) and MoSe$_2$ (blue filled squares). The gray curve is the predicted critical temperatures for MoS$_2$ in the absence of disorder, while for MoSe$_2$ the correction is negligible. The empty circles represent experimental $T_{c}$ values from Refs.\cite{doi:10.1021/acs.nanolett.8b01390,Costanzo2018,doi:10.1126/science.1228006,Saito2016,doi:10.1126/science.aab2277,AliElYumin2019}.} 
\label{fig2}
\end{figure}

The corrections to $T_c$ in the r.h.s. of Eq.\ref{eq:Tcisom} can be schematically separated in two types: the term

\begin{equation}
R_{HF}(T)/N(0)=-\dfrac{(g_1N(0)-3\lambda)}{4\pi\varepsilon_F \tau_0}(\ln \dfrac{1}{T \tau_0})^2    
\end{equation}

includes Hartree-Fock-type corrections $\propto\ln(1/T)^2$, while 

\begin{equation}
    R_v(T)/N(0) = -\dfrac{(g_1N(0)+\lambda)}{6\pi\varepsilon_F\tau_0}(\ln\dfrac{1}{T\tau_0})^3
\end{equation}

originates from  vertex corrections $\propto\ln(1/T)^3$, and constitutes the dominant correction term to the lowest order. The quantities $\lambda$, $\varepsilon_F$ and $T_{c0}$ can all be obtained through first principles calculations in the clean limit. Here we are employing  data from our previous work \cite{Marini_2023} where we have shown that calculations accounting for the sample thickness, anharmonicity and the effective field-effect geometry, leads to values of $\lambda$ in excellent agreement with experiments.  The clean limit results for $T_{c0}$ for the case of single gated MoSe$_2$ and MoS$_2$ bilayers, the minimal thickness representative of thicker flakes were superconductivity has been detected (see Ref. \cite{Marini_2023}), are compared with experimental data  in Fig. \ref{fig2}. As shown in Fig. \ref{fig2}, the clean limit correctly reproduces the superconducting onset, but $T_{c0}$ increases without saturating and no superconducting dome occurs.

The scattering time $\tau_0$ is the only quantity in Eq. \ref{eq:Tcisom} that cannot be directly estimated from a calculations on periodic systems, as it requires the inclusion of disorder. In order to evaluate the average scattering time,  we introduce a tight binding model based on maximally localized Wannier functions (MLWF)\cite{MOSTOFI2008685} having the following form in second quantization:
\begin{widetext}
\begin{equation}
H_{TB} = H_{sp} + H_{imp} = \sum_{\mathbf{R}_i\mathbf{R}_j} \sum_{l~m}
        H_{lm}(\mathbf{R}_i-\mathbf{R}_j) \,
        d_{\mathbf{R}_il}^{\dagger} d_{\mathbf{R}_jm}
     \sum_{\mathbf{R}_il} U_{l}(\mathbf{R}_i) d_{\mathbf{R}_i,l}^{\dagger} d_{\mathbf{R}_il},
  \end{equation}
  \end{widetext}
  where $d_{\mathbf{R}_il}$ and $d_{\mathbf{R}_i,l}^{\dagger} $ destroys or creates and electron in the Wannier function $l$ of the supercell $\mathbf{R}_i$. The single particle Hamiltonian $H_{sp}$ is obtained by using  MLWF and  PythTB\cite{https://doi.org/10.5281/zenodo.12721315} retaining all hopping terms larger than 2 meV (details are given in the SM, Sections S1 and S8). The diagonalization of  $H_{sp}$ leads to an electronic structure practically indistinguishable from the first principles one in the energy region relevant for superconductivity. The term $H_{imp}$ accounts for the interaction with charged impurities formed by freezing the electrolyte \cite{doi:10.1073/pnas.1716781115} and it is assumed to be diagonal in the Wannier function basis. For large values of $U_l$, the model leads to an Anderson localization. The values of $U_{l}(\mathbf{R}_i)$ are obtained by assuming  a certain density of disordered point charges generating a potential $\phi(\mathbf{r})$  on each cell of the tight binding model. The potential  $\phi(\mathbf{r})$  is calculated explicitly in the SM, Section S2 and includes both the bare potential of the external disordered charges as well as  the screening effect due to the induced charge at the Thomas-Fermi level. We thus define the on-site term $U_l$ on the site $l$ as the spatial average around each Wannier center 
\begin{equation}
    U_{l}(\mathbf{R}) = e \int d\mathbf{r}~ \phi(\mathbf{r}) |w_{l,\mathbf{R}}(\mathbf{r})|^2
\end{equation}
where $w_{l,\mathbf{R}}(\mathbf{r})$ is the $l^{th}$ Wannier function in the cell indicated by the lattice vector $\mathbf{R}$.

The physics governing the formation of the electric double layer at the interface is inherently complex\cite{Fedorov2014}. To make the problem tractable, we adopt a simplified model in which the potential $\phi(\mathbf{r})$ is computed by assuming a fixed vertical separation between the positively charged ionic impurities and the TMD layer. A lower bound to the vertical separation $d$ between the nitrogen center (approximating the charge centroid) of the DEME$^+$ cation and the topmost chalcogen layer (S or Se) can be realistically estimated based on the steric limit of the DEME$^+$ molecular geometry. We assume the cation orients its smallest substituent, the methyl group ($-\text{CH}_3$), towards the interface to minimize electrostatic energy. The distance is thus approximated as the sum of the van der Waals radius of the surface chalcogen ($r_{\text{S}} \approx 1.8~\text{\AA}$; $r_{\text{Se}} \approx 1.9~\text{\AA}$), the N--C bond length ($d_{\text{N-C}} \approx 1.47~\text{\AA}$), and the hydrogen van der Waals radius ($\approx$~1.2  \AA), which defines the steric contact distance; the tetrahedral C–H geometry reduces the vertical bond projection but does not alter the hard-sphere limit. This yields a charge-to-surface distance of $d \approx 4.6 - 4.8~\text{\AA}$, a realistic value when compared to the Stern layer thickness in ionic-liquid-gated 2D materials\cite{Jurado2017}.  We shall adopt a slightly larger value of $d~\approx$~5  \AA~ for MoS$2$ and $d\approx$~5.1  \AA~ for MoSe$2$. We make the simplest possible assumption for the impurity concentration dependence on doping, $i.e.$ that the impurity concentration $n_{imp}$ is approximately equal to the doping concentration $n_{el}$ induced by the gate potential $V$, namely we assume
  \begin{equation}
n_{imp}(V) = n_{el}(V)
\end{equation}
 which is justified if  crowding effects are neglected\cite{Fedorov2014}. With these choices, we obtain scattering time $\tau_0$ values that are compatible with the ones measured in Ref.\cite{Saito2016} at large doping for single-side gated MoS$_2$.
With these assumptions, we diagonalize the Hamiltonian, obtain eigenvalues and eigenvectors and calculate the sheet resistivity
$R_{\square}=\lim_{\omega\rightarrow0}\dfrac{1}{\sigma(\omega)d}$ where $d\approx6$ \AA ~  is the electron gas' thickness and the conductivity $\sigma(\omega)$
is calculated employing the Kubo formula \cite{PhysRevB.66.205105} as the in-plane average:
\begin{equation}
    \sigma(\omega)= <-\dfrac{2 \pi}{\Omega \omega}\sum_{n n'}|\bra{n}j_x\ket{n'}|^2 (f_n-f_{n'})\delta(\hbar\omega-\varepsilon_{n'}+\varepsilon_{n})>\label{eq:sigma}
\end{equation}
where $\Omega$ is the supercell volume and $f_n$ is the occupation of the eigenstate $|n\rangle$ and $x$ is a generic in-plane direction. The $\alpha$ cartesian components of electric current density operator are obtained as \[ j_{\alpha}=-\frac{e i}{\hbar} \sum_{l~m} H_{lm}(\mathbf{R}_i-\mathbf{R}_j) \,(\mathbf{R}_i-\mathbf{R}_j)_\alpha  d_{\mathbf{R}_il}^{\dagger} d_{\mathbf{R}_jm}\].
In practical calculations, a finite broadening $\eta$ is added to calculate the Dirac $\delta$ function and $\tau_0$ can be obtained by performing two different calculations at the same $\eta$, one in the presence of disorder (giving $R_{\square}^\eta$) and one in the absence of disorder (giving $R_{\square}^{\eta,0}$) . The scattering time in the absence of $\eta$ is then obtained as $\tau_0=\dfrac{2}{\eta} (R_{\square}^\eta/R_{\square}^{\eta,0}-1)^{-1}$ (see the SM, Section S1). The obtained two-dimensional conductivity $\sigma$ for a MoS$_2$ gated bilayer as a function of impurity concentration is reported in the SM, Section S1. As expected, disorder strongly suppresses the Drude peak, in agreement with mobility measurements\cite{doi:10.1073/pnas.1716781115,Saito2016,doi:10.1126/science.1228006}. The calculated $\tau_0$, Fermi level $\varepsilon_F$, sheet resistivity $R_{\square}$ for a gated  MoS$_2$ bilayer are reported in table \ref{tab1m}.

\begin{table}[htp]
    \centering
    \begin{tabular}{|c|c|c|c|c|c|}
         \hline
  system  & n$_e$/cell  & R$^0_{\square}(\Omega)$ &  $\varepsilon_F^{avg}$ & $\tau_0$ (fs)  &  $R_{HF+v}(T_c)/N(0)$  \\
         \hline 
                  \hline
          bilayer & 0.1 &  52.49  & 0.05 & 169.85 & -0.064 \\
\hline
          bilayer & 0.125 &  82.49 & 0.06 & 108.08 & -0.135 \\
          \hline
           bilayer & 0.175  & 139.71 & 0.08 & 50.56 & -0.733 \\
         \hline
    \end{tabular}
    \caption{Model-derived parameters entering the self consistent formula for $T_{c}$ and the equation of the superconducting gap in the presence of disorder.} \label{tab1m}
    \end{table}

    The self-consistent solution of Eq. \ref{eq:Tcisom} is displayed in Fig. \ref{fig2} for gated MoS$_2$ and MoSe$_2$ bilayers. A superconducting dome appears in bilayer MoS$_2$ in qualitative and quantitative agreement with experiments. At carriers concentration of $n_e=2\times 10^{14} ~$e$^-/$cm$^2$ the $T_{c0}$ suppression due to disorder is approximately $300\%$. Remarkably, even using the same disorder parameters, a negligible suppression of the critical temperature in the case of a gated MoSe$_2$ bilayer  
is observed, at least up to 1.69 $e^-\times$ 10$^{14}~e^-$/cm$^{2}$\cite{PMID:26235962}, as shown in Fig.\ref{fig2}. The reason is that the minimum at K is never occupied in this case \cite{Marini_2023}, generating a higher Fermi energy in the Q-valley (a higher carrier velocity at comparable doping), and thus making the disorder $T_{c}$ renormalization less relevant. This is investigated in more detail in the SM, Section S5, where we discuss the behavior of $T_{c}$ as a function of $\tau_0$ and $\varepsilon_F$ for a 2D homogeneous electron gas.

Following our analysis, the superconducting dome in gated TMDs is due to two competing effects. The onset of the superconducting dome is determined by the occupation of the $Q$ valleys switching on the electron-phonon interaction that grows proportionally to the Fermi level. The saturation of $T_{c}$ and the consequent occurrence of a superconducting dome at high voltage is determined by the disorder increasing at higher voltages. We speculate that disorder will saturate at even larger doping values due to crowding effects\cite{Fedorov2014}, leading to a finite $T_{c}$ even at very large doping (see also Fig. S8 in the SM).

\begin{figure}[t!]
\centerline{\includegraphics[width=0.9\columnwidth]{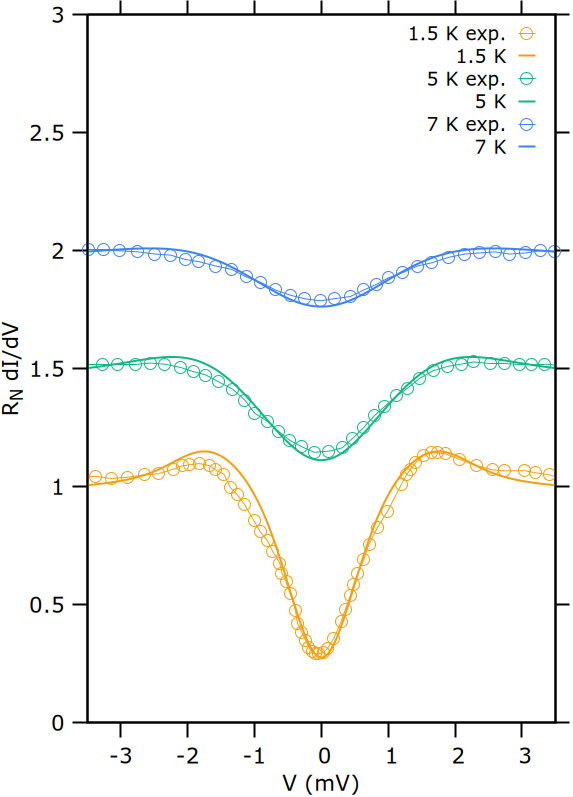}}
\caption{Theoretical versus experimental differential tunneling conductance for single-side gated MoS$_2$ at $n_e \approx 1.5 \times 10^{-14}/\mathrm{cm}^2$ as a function of voltage at 1.5 (orange) 5 (green) and 7 (blue) Kelvin. Experimental data have been extracted from Ref.\cite{Costanzo2018}. } 
\label{fig3}
\end{figure}

While the above analysis can accurately reproduce the $T_{c}$ versus voltage phase diagram of dichalcogenides, an appropriate description of the superconducting
state must be also able to account for the experimentally measured superconducting gap. In particular, the superconducting gap has a distinctive V-shape \cite{Costanzo2018} that is not compatible with a standard s-wave phonon-mediated pairing in the clean limit, as the gap would be substantially larger and no V-shape would occur \cite{Marini_2023}. We now show that within our framework explicitly including disorder we obtain
an extremely accurate description of the superconducting gap and its dependence on temperature.

The modified gap equation in the presence of homogeneous disorder reads:
\begin{equation}
    1=V^{BCS}\left[\sum_n^{N} \frac{\tanh(\dfrac{E_n}{2T})}{E_n}+\tilde{R}_{HF}+\tilde{R}_v\right]\label{eq:multigap_main}
\end{equation}
where $V^{BCS}$ is the BCS potential and we assumed a single band for simplicity, while $\tilde{R}_{HF}$ and $\tilde{R}_v$ are disorder-induced corrections in the superconducting state, defined the SM Sec. S3, where we give the full derivation and a generalization to the multi-band case. In the isotropic approximation, this equation leads to an exponential suppression of the gap at $T=0$ K with respect to the disorder parameter $x=-1/N(0)(\tilde{R}_{HF}+\tilde{R}_v)$ :
\begin{equation}
\Delta_0(x) \sim \theta_D ~e^{-1/\lambda}e^{-x}   \label{eq:deltam} 
\end{equation}
where $\theta_D$ is the Debye frequency. 
As it is reasonable to speculate that disorder is horizontally inhomogeneous across the the sample, $x$ does not takes a single value but it can vary from point to point. For simplicity, we assume that $x$ is uniformly distributed in an interval $[a,b]$ where $a$ is the value leading to the calculated $T_{c}$ in Fig. \ref{fig2} ($\approx 0.2$ at the experimental doping), while $b$ is a much larger value, which we fix to 2.3. We then solve the self-consistent gap equation for many values of $x$ and calculate the tunneling conductance defined as\cite{Tinkham1975IntroductionTS}:\\

\begin{equation}
    R_N\dfrac{dI}{dV}(V) = -\dfrac{1}{e^-}\int_{-\infty}^{\infty}d\varepsilon \dfrac{N_S(\varepsilon)}{N(0)} \cdot\dfrac{\partial{f}}{{\partial V}}(\varepsilon+e^-V,T)
    \label{eq:cond}
\end{equation}

where $f$ is the Fermi-Dirac distribution, $N_S(\varepsilon)$ is the superconducting density of states, defined as:

\begin{equation}
    N_S(\varepsilon)= N(0)~\mathrm{Re}\left\{\dfrac{|\varepsilon+i\Gamma|}{\sqrt{(\varepsilon+i\Gamma)^2-\Delta^2}}\right\}\label{eq:sdos}
\end{equation}

and R$_N$ is the normal state resistance. In Eq.\ref{eq:sdos}, $\Gamma$ is the quasiparticle linewidth $\Gamma$\cite{PhysRevLett.41.1509}, which we set to an infinitesimal value, $\Gamma=0.01$~meV, without the necessity to include an unphysical quasiparticle linewidth to fit the tunneling spectra (see the discussion in the SM of Ref.\cite{Costanzo2018}) and the SM of the present paper, Secs. 3,8). 

Our results on the superconducting gap calculations in the presence of disorder are validated against experiments in Fig. \ref{fig3} for MoS$_2$ thick ($> 1$ layer) samples.
Our calculations are in excellent agreement with experiments and explain, (i) the reduction of the superconducting gap with respect to the clean limit (see Fig. 22 in \cite{Marini_2023}), (ii) the V-shape feature, (iii) the presence of a kink at approximately $\pm 1$ mV in the $1.5$~ K tunneling spectra and, finally, (iv) the temperature dependence of the tunneling spectra. In the present scenario, the V-shape feature is not a signature of unconventional pairing symmetry, rather it reflects the action of non-homogeneous disorder in proximity of an Anderson transition on a BCS gap. Multi-band effects are negligible in these systems as shown in \cite{Marini_2023}, nevertheless they give rise to interesting considerations, exposed in Sec. S3 in the SM.
Within the current framework, we are able to account for the magnitude, the shape, and the temperature dependence of the superconducting gap.

In conclusion, we demonstrate that strong electron-phonon interaction and disorder enhancement of the Coulomb interaction are the key mechanism governing the shape of the superconducting dome in ionic-liquid-gated MoS$_2$.  Our 
results for the superconducting critical temperature versus voltage in MoS$_2$ and MoSe$_2$ are in excellent agreement with experimental data and represents qualitative improvement with respect to all other attempts to explain the phase diagram of gated dichalcogenides.
Finally, by extending disorder corrections into the superconducting state via a modified BCS framework further allows us to account for the kink observed in tunneling spectra and to reproduce its characteristic V-shape.\\
Our results establish disorder-driven enhancement of Coulomb repulsion due to the the proximity of an Anderson transition as a universal feature of ionic-liquid-gated quasi-two-dimensional materials subject to high bias. Our joint many-body and first principles framework provides a unified interpretation of dome superconductivity and tunneling spectra in gated dichalcogenides, offering new insight into the microscopic origin of superconducting dome physics. In the future, the same approach can be applied to any other gated quasi-two-dimensional material following the same procedure adopted in this paper.\\

We acknowledge the EuroHPC Joint Undertaking for awarding this project access to the EuroHPC supercomputer LEONARDO, hosted by CINECA (Italy) and the LEONARDO consortium through an EuroHPC Regular Access call. We acknowledge ISCRA (ISCRA B EPhoCS) for awarding this project access to the LEONARDO supercomputer, owned by the EuroHPC Joint Undertaking, hosted by CINECA (Italy). This work was funded by the European Union (ERC, DELIGHT, 101052708). Views and opinions expressed are however those of the author(s) only and do not necessarily reflect those of the European Union or the European Research Council. Neither the European Union nor the granting authority can be held responsible for them.

All data generated to obtain the results presented in the paper will be made available in a zenodo repository upon publication.

\clearpage
\onecolumngrid

\renewcommand{\thepage}{\arabic{page}}
\renewcommand{\thesection}{S\arabic{section}}
\renewcommand{\thetable}{S\arabic{table}}
\renewcommand{\thefigure}{S\arabic{figure}}

\makeatletter

\newcommand*{\addFileDependency}[1]{% argument=file name and extension
\typeout{(#1)}% latexmk will find this if $recorder=0
% however, in that case, it will ignore #1 if it is a .aux or 
% .pdf file etc and it exists! If it doesn't exist, it will appear 
% in the list of dependents regardless)
%
% Write the following if you want it to appear in \listfiles 
% --- although not really necessary and latexmk doesn't use this
%
\@addtofilelist{#1}
%
% latexmk will find this message if #1 doesn't exist (yet)
\IfFileExists{#1}{}{\typeout{No file #1.}}
}\makeatother

\newcommand*{\myexternaldocument}[1]{%
\externaldocument[MAIN-]{#1}%
\addFileDependency{#1.tex}%
\addFileDependency{#1.aux}%
}
%\myexternaldocument{main}
%\externaldocument{prx_si}

%\begin{document}
\preprint{APS}

\begin{center}
{\Large\bf Supplemental Material for "Disorder-Driven Enhancement of Coulomb Repulsion Governs The Superconducting Dome in Ionic-Liquid-Gated Quasi-2D Materials"}\\[0.5em]

Giovanni Marini$^{1,2}$, Pierluigi Cudazzo$^1$ and Matteo Calandra$^{1,2}$\\
\textit{1 Department of Physics, University of Trento, Via Sommarive 14, 38123 Povo, Italy}\\
\textit{2 Graphene Labs, Fondazione Istituto Italiano di Tecnologia, Via Morego, I-16163 Genova, Italy}
\end{center}

%\title{Machine learning for large-scale long-time simulations of light-induced order-disorder phase transitions with {\it ab initio} accuracy.}

%\author{Giovanni Marini}
%\affiliation{Department of Physics, University of Trento, Via Sommarive 14, 38123 Povo, Italy}

%\author{Pierluigi Cudazzo}
%\affiliation{Department of Physics, University of Trento, Via Sommarive 14, 38123 Povo, Italy}

%\author{Matteo Calandra} 
%\affiliation{Department of Physics, University of Trento, Via Sommarive 14, 38123 Povo, Italy}

%\maketitle

\tableofcontents

\section{Calculation of $R_\square$ and scattering time $\tau_0$ in a Wannier tight binding approach}
\label{app:a}
\begin{figure}[t]
\includegraphics[width=0.5\columnwidth]{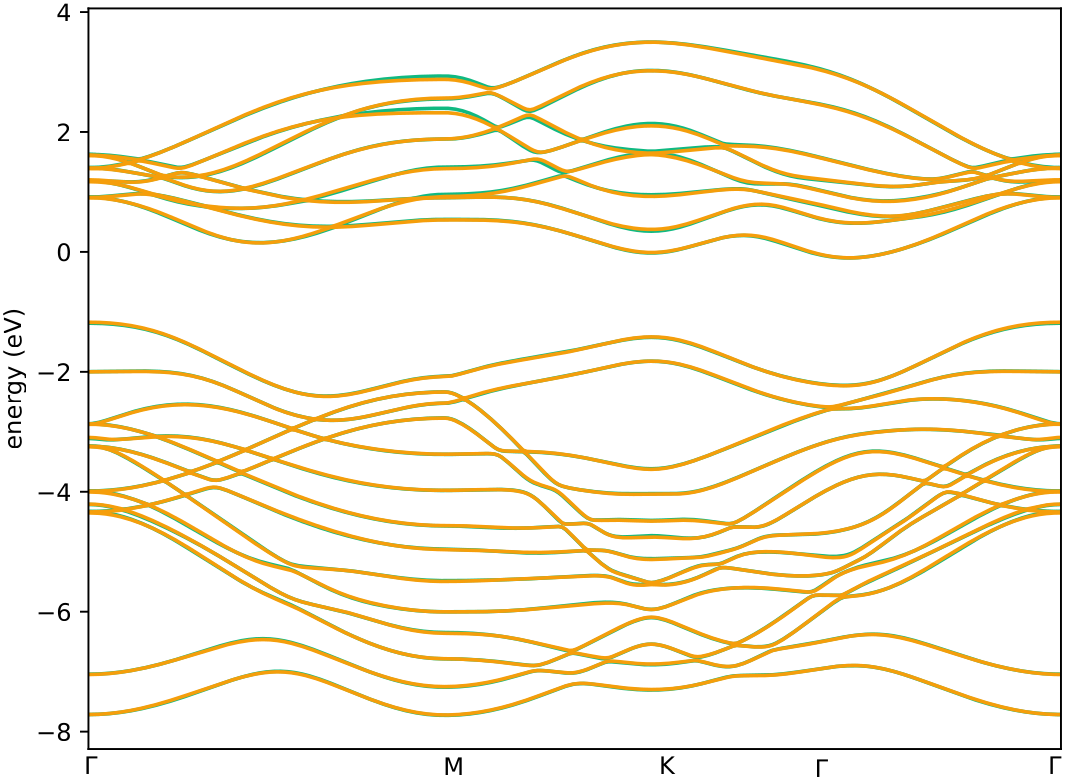}
\caption{Electronic band structure comparison between PythTB (orange) and Wannier90 (green) for bilayer MoS$_2$ at $n_e = 0.175~$/cell.} 
\label{figS4}
\end{figure}

We simulate transport properties of gated monolayer and bilayer MoS$_2$ from first principles starting from density-functional theory calculations in the gate geometry, according to the methodology presented in Ref.\cite{PhysRevB.96.075448}. Density-functional theory calculations are performed employing the same prescriptions described in Ref.\cite{Marini_2023} employing Quantum ESPRESSO\cite{Giannozzi_2009}. In order to efficiently include disorder, we describe the system in the Wannier basis employing Wannier90\cite{MOSTOFI2008685} for the Wannierization. The Wannier model is composed by 22 (11 for spin) states for the monolayer (doubled for the bilayer), 10 $d$ states for the molybednum and 6 $p$ states for every sulfur in the unit cell. 
 The tight-binding Hamiltonian is extracted employing PythTB\cite{https://doi.org/10.5281/zenodo.12721315} and retaining all hopping terms larger than 2 meV. The quality of the resulting band structure can be checked in Fig.\ref{figS4}, where we compare the first principles band structure obtained from Quantum ESPRESSO and PythTB. In real space, the Wannier Hamiltonian is fully characterized by the matrix elements $H_{ij}(\mathbf{R})$:

\begin{equation}
    H_{lm}(\mathbf{R})=\bra{\phi_{0l}}\hat{H}\ket{\phi_{\mathbf{R}m}}
\end{equation}

where $\mathbf{R}$ is a lattice vector and $l$ and $m$ are two Wannier function indices. A $k$-space Hamiltonian is then built assuming periodicity as a Bloch sum of these matrix elements\cite{https://doi.org/10.5281/zenodo.12721315}:

\begin{equation}
    H_{lm}({\mathbf{k}})=\sum_{\mathbf{R}}e^{i \mathbf{k}\cdot{(\mathbf{R}+\mathbf{\tau}_m-\mathbf{\tau}_l)}}H_{ij}(\mathbf{R}) \label{eq:Hijk}
\end{equation}

We calculate the optical conductivity starting from the standard expression for velocity operator matrix element in periodic systems\cite{10.21468/SciPostPhysCore.6.1.002,10.21468/SciPostPhys.12.2.070,Ashcroft76}:

\begin{equation}
    \bra{n\mathbf{k}}\hat{v}\ket{n'\mathbf{k}}=\sum_{lm} u^*_{l n}(\mathbf{k})u_{m n'}(\mathbf{k})\nabla_\mathbf{k}H_{lm}(\mathbf{k})
\end{equation}

where $\ket{n\mathbf{k}}$ are Bloch states and the $u_{ln}$ is the projection of the Wannier function $l$ on the usual Bloch like basis functions\cite{https://doi.org/10.5281/zenodo.12721315}.

Here, the Berry connection term in the velocity matrix element is neglected for simplicity\cite{10.21468/SciPostPhysCore.6.1.002,10.21468/SciPostPhys.12.2.070}. We then obtain the following expression for the microscopic current operator:
\begin{equation}
\bra{n\mathbf{k}}\hat{\mathbf{j}}\ket{n'\mathbf{k}} = \sum_{lm} u^*_{l n}(\mathbf{k})u_{m n'}(\mathbf{k}) \dfrac{e}{\hbar} \nabla_\mathbf{k} H_{lm}(\mathbf{k})= i\dfrac{e}{\hbar}\sum_{lm} u^*_{l n}(\mathbf{k})u_{m n'}(\mathbf{k}) (\mathbf{R}+\mathbf{\tau}_m-\mathbf{\tau}_l)H_{lm}(\mathbf{R})
\end{equation}

 Disorder is included as on-site terms in our tight binding models. Each site acquires an on-site contribution $-~U_l=H_{ll}(0)$ that depends on the position of the charged impurities, see Appendix~\ref{app:b}. In the presence of disorder, periodicity is broken and the Bloch sums are built for a large supercell, see Appendix \ref{app:e}. The sheet resistivity is $R_{\square}=\lim_{\omega\rightarrow0}\dfrac{1}{\sigma(\omega)d}$ where $d$ is the thickness and the conductivity $\sigma(\omega)$ is calculated in linear response employing the Kubo formula\cite{PhysRevB.66.205105} as an in-plane average of the quantity:

\begin{equation}
    \sigma(\omega)=<\sigma_{xx}(\omega)>_{plane}= <\dfrac{2}{\Omega \omega}\sum_{n n'}|\bra{n}j_x\ket{n'}|^2 \mathrm{Im}\big(~\dfrac{f_n-f_{n'}}{\hbar \omega-\varepsilon_{n'}+\varepsilon_n+i\eta}\big)>\label{eq:sigma0}
\end{equation}

where $\Omega$ is the supercell volume and $f_n$ is the occupation for state $n$, $\eta$ is a small broadening and $x$ is a generic in-plane direction. For small broadening $\sigma_{xx}(\omega)$ is equivalently rewritten as

\begin{equation}
    \sigma_{xx}(\omega)= -\dfrac{2 \pi}{\Omega \omega}\sum_{n n'}|\bra{n}j_x\ket{n'}|^2 (f_n-f_{n'})\delta(\hbar\omega-\varepsilon_{n'}+\varepsilon_{n})
\end{equation}

 For a periodic system, $\sigma_{xx}$ in Eq.~\ref{eq:sigma0} can be written as:

\begin{equation}
    \sigma_{xx}(\omega)= -\dfrac{2 \pi }{\Omega N_k \omega}\sum_{n n'\mathbf{k}}|\bra{n\mathbf{k}}j_x\ket{n'\mathbf{k}}|^2 (f_{n\mathbf{k}}-f_{n'\mathbf{k}})~\mathrm{Im}\big(~\dfrac{1}{\hbar \omega-\varepsilon_{\mathbf{k}n'}+\varepsilon_{\mathbf{k}n}+i\eta}\big) \
\end{equation}

In the  $\omega\to 0$ limit, the intraband (Drude) term can be evaluated employing the derivative of the Fermi function: 

\begin{equation}
    \sigma^{intra}(\omega)= <-\dfrac{2 \pi  }{\Omega N_k}\sum_{n \mathbf{k}}|\bra{n\mathbf{k}}j_x\ket{n\mathbf{k}}|^2 \dfrac{df}{d\varepsilon} \mathrm{Im}\big(~\dfrac{1}{\hbar \omega+i\eta}\big)> \
\end{equation}
In practice, we assumed in-plane isotropy and calculated the conductivity along the (100) direction. From this expression, the sheet resistivity $R_\square$ is defined as

\begin{equation}
    R_{\square}=1/\sigma_{2D}=\dfrac{1}{\sigma(0)d}
\end{equation}

where $d$ is the two-dimensional (2D) electron gas thickness and we assume isotropy in the plane. In order to extract the scattering time $\tau_0$ from a practical calculation in the presence of disorder one can operate at finite broadening $\eta$ and employ the usual semiclassical relationship:

\begin{equation}
\dfrac{1}{\tau_{\eta}} = \dfrac{1}{\tau_0}+\dfrac{1}{\tau^0_{\eta}}
   \end{equation}

%  \sigma =  e^2N(0) v_x^2 \tau_0 = e^2N(0) \dfrac{v_F^2}{2} \tau_0

where $\tau^0_{\eta}=\dfrac{1}{2\eta}$ is the scattering introduced by the broadening $\eta$, while $\tau_{\eta}$ is the total scattering time (disorder plus broadening) in the presence of a broadening $\eta$. Since  $R_{\square} \propto 1/\tau$\cite{RevModPhys.57.287,doi:10.1143/JPSJ.51.1380} one can compare (at fixed broadening) the calculated sheet resistivity in the presence of disorder $R_{\square}^{\eta}$ and the one in the absence of disorder $R_{\square}^{\eta,0}$ to find:

\begin{equation}
    \dfrac{R_{\square}^{\eta}}{R_{\square}^{\eta,0}}=\dfrac{\tau^0_{\eta}}{\tau_{\eta}}= 1 + \dfrac{\tau_{\eta}^0}{\tau_0}
\end{equation}

and from this relationship it is possible to extract $\tau_0$. From $\tau_0$, one finally finds $R_\square^0$ from the relationship:

\begin{equation}
      \dfrac{R_{\square}^{0}}{R_{\square}^{\eta}}=\dfrac{\tau_{\eta}}{\tau_0}
\end{equation}

%%%%%%%%%%%%%%%%%%%%%%

The obtained two-dimensional conductivity $\sigma_{2D}(\omega)$ is reported in Fig.\ref{figS0} (panel a) for single-side gated bilayer MoS$_2$ for selected impurity concentration values, showing a systematic low-frequency suppression with increasing disorder. The same conclusion holds also in the monolayer and double gated system: in panel b) of Fig.\ref{figS0} we study the static conductivity $\sigma_{2D}(0)$ for four exemplificative cases:

\begin{figure}
\includegraphics[width=1.0\columnwidth]{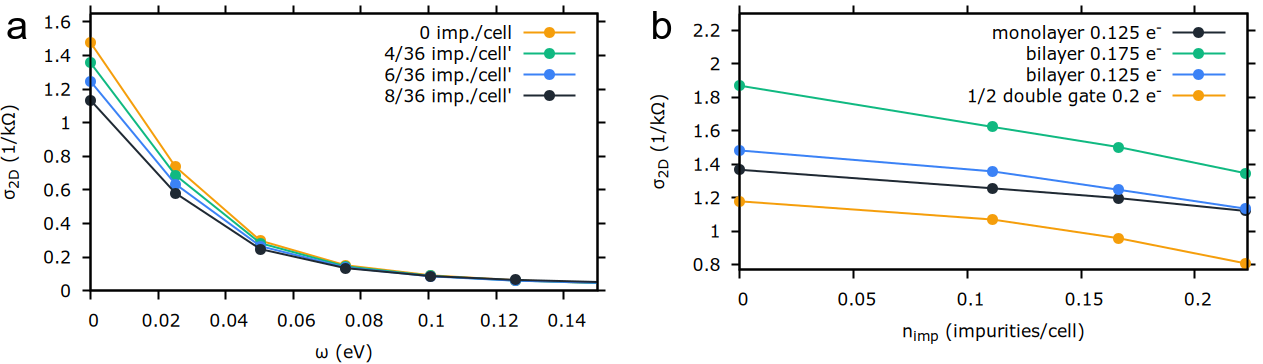}
\caption{Panel a):calculated two-dimensional conductivity $\sigma_{2D}(\omega)$ with a broadening $\eta=0.025$~eV for single side gated bilayer MoS$_2$ at $n_e = 0.125 ~e^-$/cell for various impurity concentration values $n_{imp}$. Panel b): two-dimensional conductivity $\sigma_{2D}$ as a function of the impurity concentration values $n_{imp}$ for selected systems (double gate conductivity is divided by two for graphical reasons). } 
\label{figS0}
\end{figure}

\begin{enumerate}
\item Single side gated monolayer MoS$_2$ 
at $n_e =~0.125~e^-$/cell. 
\item Single side gated bilayer MoS$_2$
at $n_e =~0.125~e^-$/cell.
\item Single side gated bilayer MoS$_2$ 
at $n_e =~0.175~e^-$/cell.
\item Double side gated bilayer MoS$_2$ 
at $n_e =~0.2~e^-$/cell.
\end{enumerate}

We observe an universal suppression of the static conductivity with increasing impurity concentration, as expected.  Contrary to what one may expect, we do not observe a more marked suppression in the monolayer. The double side gated MoS$_2$ presents a much larger initial conductivity. However, disorder effects on superconductivity for this system and the absence of a dome feature in the experiments\cite{Zheliuk2019}~cannot be studied with the present approach, since the double gate system cannot be  modeled by a single layer of charge and strongly differs from the other studied cases, thus requiring to take into account the quasi-2D nature of the electron gas.
%%%%%%%%%%%%%%

Finally, it is worth recalling that in a purely two-dimensional system every state is localized and the conduction is non-ohmic\cite{PhysRevLett.42.673}, at least in the absence of spin-orbit coupling and arbitrarily strong interactions\cite{RevModPhys.57.287,ALTSHULER1983429}, where the situation is more nuanced. The localization length can be arbitrarily large\cite{RevModPhys.57.287} and the conductance for a sample of linear dimension $L$, $g(L)$, can be written as:

\begin{equation}
    g(L)=g_0-g_1(L)
\end{equation}

where $g_0$ is the microscopic (Boltzmann) dimensionless conductance and $g_1$ contains model-dependent renormalization. For example, in the case of a spinless two-dimensional system one has\cite{RevModPhys.57.287}:

\begin{equation}
    g(L)=g_0 - \dfrac{e^2}{\hbar \pi^2} \ln(\dfrac{L}{l})
\end{equation}

These system-dependent considerations contribute to determine the scale dependent corrections to the sheet resistivity $R_\square$ measured experimentally. The theoretical $R_\square=\dfrac{e^2}{h g_0}=1/\sigma_{2D}$ calculated in our model is the microscopic one.

\section{Modeling of charged impurities and screened potential}
\label{app:b}

In this Appendix~we discuss how the effects of external impurity charges were included in the tight binding model. In the case of interest here, we want to simulate the external potential acting on the material due to the free ions in the liquid gate. In general, this is a disordered potential produced by the arrangement of the free ions. A straightforward way to proceed is to include the potential self-consistently in the density-functional self-consistent calculation. This approach would however be very computationally intensive due to the large supercells required to simulate the disorder. In order to overcome this difficulty we will assume that, to a first approximation, the external charge distribution is well represented by an ordered and homogeneous plane of charge, which we include in the self consistent calculation via the approach discussed in Ref.\cite{PhysRevB.96.075448}, plus a certain density of disordered point charges that induce a disordered potential on the quasi two-dimensional material. Our aim is to calculate the disorder induced potential on each lattice site of the tight binding model. To do this, a straightforward way is to perform a spatial average around each Wannier center and define the on-site term $U_i$ on the site $i$ as

\begin{equation}
    U_{i}(\mathbf{R}) = e \int d\mathbf{r}~ \phi(\mathbf{r}) |w_{i,\mathbf{R}}(\mathbf{r})|^2 \label{eq:int}
\end{equation}

where $w_{i,\mathbf{R}}(\mathbf{r})$ is the $i^{th}$ Wannier function in the cell indicated by lattice vector $\mathbf{R}$. To calculate the potential $\phi(\mathbf{r})$ one needs to consider both the potential from the external disordered charges and the screening effect due to the induced charge. The starting point is the Poisson's equation, which relates an external electrostatic potential $\phi$ to the total charge density $\rho$ as:

\begin{equation}
    \nabla\cdot (\kappa(\mathbf{r}) \nabla \phi(\mathbf{r})) = - 4 \pi \rho(\bf{r}) \label{eq:poisson}
\end{equation}

here $\kappa$ represents the (generally position dependent) relative permittivity and $\rho$ is the total charge. Note that we are using Gaussian units in this derivation. For simplicity, we will consider a position-independent $\kappa$.
The charge density is split in two contributions, $\rho(\mathbf{r}) = \rho_{ext}(\mathbf{r})+ \rho_{ind}(\mathbf{r})$, where $\rho_{ext}(\mathbf{r})$ represents the external charge and $\rho_{ind}(\mathbf{r})$ represents the induced charge on the 2D material. Since the gated 2D material hosts free charges, the screening will be of the metallic type. We shall assume a local response in the Thomas-Fermi spirit, first discussed by Stern and Howard\cite{PhysRev.163.816,RevModPhys.54.437} for the case of two-dimensional systems. To first order in the field one assumes that the eigenfunctions $\psi_i(\mathbf{r})$ are not changed and that the induced charge results from the local variation of the chemical potential:

\begin{equation}
    \rho_{ind}(\mathbf{r}) = - e~[n(\varepsilon_F+e^-\phi(\mathbf{r}))-n(\varepsilon_F)]
\end{equation}

where $n(\mathbf{r},\varepsilon_F)$ represents the charge density in the material. Assuming a small $\phi(\mathbf{r})$ one can linearize this equation and obtain the usual linear Thomas-Fermi expression:

\begin{equation}
\rho_{ind}(\mathbf{r})  = -e^2 \dfrac{\partial n(\varepsilon_F)}{\partial \varepsilon_{F}}\phi(\mathbf{r})
\label{eq:rhoind}
\end{equation}

We now focus on the external charge. In the case of interest, this is a positively charged DEME$^+$ molecule, which we represent as a $\delta$ charge found at a certain $z_0<0$ (for definiteness, we establish that the 2D material's out-of-plane thickness extends in the positive $z$ direction and the ionic liquid gate in the negative $z$ direction).

\begin{figure}[t]
\includegraphics[width=0.45\columnwidth]{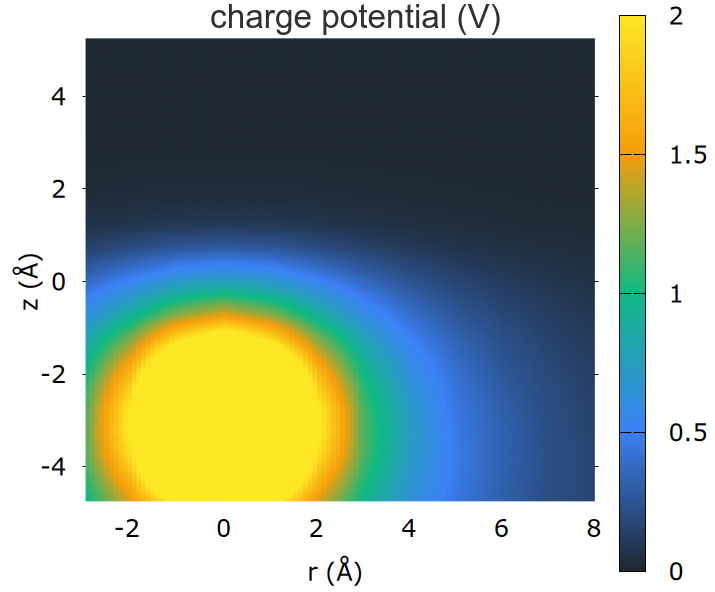}
\caption{Charge potential induced by a positive charge placed in $(r,z_0)=(0,-3.0$) \AA. The chalcogen nuclei in the 2D material are at $z=2$ \AA.} 
\label{figS5}
\end{figure}

For this particular choice of the external charge, we assume that the potential $\phi(\mathbf{r})$ only depends on the radial distance from the impurity $r_2 = (x^2+y^2)^{1/2}$ and from the out of plane distance $z$, $\phi(\mathbf{r}) = \phi(r_2,z)$.

We further simplify the problem by assuming that the  derivative of the charge density can be written as

\begin{equation}
    \dfrac{\partial n(\varepsilon_F)}{\partial \varepsilon_F} = \dfrac{\partial n_{2D}(\varepsilon_F)}{\partial \varepsilon_F} g(z)
\end{equation}
, where we introduced the two-dimensional charge density $n_{2D}(r_2,\varepsilon_F)$ and an envelope function $g(z)$, which is normalized to one in $z$. This is a reasonable approximation since the doping charge is almost all localized on top of a single sheet of atoms, the one closest to the ionic liquid\cite{Marini_2023}. We adopt the following analytic common choice for the enevelop function 

\begin{equation}
    g(z) = \frac12b^3z^2e^{-bz}
\end{equation}

 where $b$ is a dimensional parameter which we set to~1 \AA$^{-1}$~, fitted to the extension of the induced first-principles charge showed in Ref.\cite{Marini_2023}, with the understanding that $\int_0^\infty g(z)=1$. We also verified that using the real induced charge profile doesn't alter the results significantly. 
Substituting the expressions for $\rho_{ind}$ from Eq.\ref{eq:rhoind} and the $\delta$ form for $\rho_{ext}$ in Eq.~\ref{eq:poisson} one obtains
\begin{equation}
     \nabla^2 \phi(r_2,z) -2 q_s g(z)\phi(r_2,z) = - 4 \pi/\kappa~ \delta(z-z_0)\delta(\mathbf{r}_2)\label{eq:poisson2}
\end{equation}

 where we introduced the Thomas-Fermi screening parameter ${q}_{s}$ defined by: 

\begin{equation}
    q_s = \dfrac{2 \pi e^2}{\kappa} \dfrac{\partial n_{2D}(\varepsilon_F)}{\partial \varepsilon_F}
\end{equation}

Eq.~\ref{eq:poisson2} has no easy analytical solution but can be solved numerically (see Appendix \ref{app:e}). In Fig.\ref{figS5} we plot the electrostatic potential induced by a free positive charge in $z_0=-3.0$ \AA~ as a function of the radial distance $r$ and the vertical distance $z$, assuming a Thomas-Fermi screening vector $q_s = 8.35 $~\AA$^{-1}$. At $z=0$, where the outer layer of charge of the system is expected, we report potential values in the range that never exceed 0.8 eV, the precise value depending on the radial distance from the charge.
In order to gain further insights from the analytical treatment, we make an additional simplification following Ref.\cite{PhysRev.163.816}, assuming that the potential felt by the induced charge is the average potential in z: 

\begin{equation}
    {\phi}_{avg}(r_2) = \int_0^\infty dz~\phi(r_2,z)g(z)
\end{equation}

Eq. \ref{eq:poisson2} then becomes: 

\begin{equation}
     \nabla^2 \phi(r_2,z) -2 q_s g(z){\phi}_{avg}(r_2) = - 4 \pi/\kappa~ \delta(z-z_0)\delta(\mathbf{r}_2)\label{eq:poisson3}
\end{equation}

We are interested in the solution in a generic point in the material, $(r,z\geq0)$. The way to proceed is by express the potential using the Hankel transform\cite{PhysRev.163.816,RevModPhys.54.437}:

\begin{equation}
    \phi(r_2,z) = \int_0^\infty dq~q~\tilde{\phi}(q,z)J_0(qr_2) 
\end{equation}

where $J_0$ is the $0^{th}$ order Bessel function of the first kind.
In this way the following equation for $\tilde{\phi}$ is obtained:

\begin{equation}
    (\partial_z^2-q^2)\tilde{\phi}(q,z)-2 q_s \tilde{\phi}_{avg}(q)g(z)=-2e\delta(z-z_0)
    \label{eq:henkspace}
\end{equation}

The Green's function $G_q(z,z')$ for the modified Helmoltz operator is:

\begin{equation}
\begin{gathered}
    (\partial_z^2-q^2) G_q(z,z') = -\delta(z-z')\\
    G_q(z,z') = \dfrac{e^{-{q|z-z'|}}}{2q}
\end{gathered}
\end{equation}

and it is straightforward to check that the general solution can be expressed in terms of $G_q(z,z')$ as:

\begin{equation}
\begin{gathered}
        \tilde{\phi}(q,z) = \dfrac{2e G_q(z,z_0)}{\kappa}-2q_s\tilde{\phi}_{avg}(q)\int_0^\infty dz' G_q(z,z')g(z') \\
        = \tilde{\phi}_0(q,z) + \tilde{\phi}_1(q,z)
\end{gathered}
\end{equation}

and an expression for $\tilde{\phi}_{avg}$ is obtained integrating the solution on both sides:

\begin{equation}
    \tilde{\phi}_{avg}(q) = \tilde{\phi}^0_{avg}(q)/[1+q_s K(q)]
\end{equation}

where we introduced the Kernel 
\begin{equation}
   K(q) = \int_0^\infty dz \int_0^\infty dz'~ g(z) ~g(z')~ 2G_q(z,z') 
\end{equation}

It is straightforward to verify that one recovers the usual 2D Thomas-Fermi for $g(z) = \delta(z)$. 

The model can be straightforwardly extended to the case where the charge is distributed over more than a single layer, by expressing the envelop function as a sum over layers:

\begin{equation}
        g(z) = \sum_{i=1}^N a_i g_i(z)= \sum_{i=1}^N~a_i \frac12b_i^3(z-z_i)^2e^{-b_i(z-z_i)}
\end{equation}

Here, $z_1=0$ and the $z_{i+1}-z_i$ is the interlayer distance, with the usual understanding that $\int_0^\infty g(z)=1$ and $\sum_{i=1}^N a_i= 1$. The average potential felt by the charge on the $i^{th}$ layer is: 

\begin{equation}
    \phi^i_{avg}(r_2)=\int_0^\infty dz ~\phi(r_2,z) g_i(z)
\end{equation}

where the denominator serves as a normalization. In this case, Eq.~\ref{eq:poisson2} is simplified as: 

\begin{equation}
      \nabla^2 \phi(r_2,z) -2 q_s \sum_{i=1}^{N}~a_ig_i(z)\phi^i_{avg}(r_2) = - 4 \pi/\kappa~ \delta(z-z_0)\delta(\mathbf{r}_2)\label{eq:poisson4}
\end{equation}

Applying the Hankel transform and obtain an equation analogous to Eq.\Ref{eq:henkspace}:

\begin{equation}
    (\partial_z^2-q^2)\tilde{\phi}(q,z)-2 q_s \sum_{i=1}^N a_i~\tilde{\phi}^i_{avg}(q)g_i(z)=-2e\delta(z-z_0) \label{eq:henk2}
\end{equation}

In this case the general solution can be expressed as:

\begin{equation}
            \tilde{\phi}(q,z) = \dfrac{2e G_q(z,z_0)}{\kappa}-2q_s\sum_{i=1}^N~a_i\tilde{\phi}^i_{avg}(q)\int_0^\infty dz' G_q(z,z')g_i(z')  = \tilde{\phi}_0(q,z)+\sum_{i=1}^N \tilde{\phi}_1(q,z) \label{eq:henkm}
\end{equation}

and the expression for $\phi^i_{avg}$ is again obtained multiplying left and right members of Eq.\ref{eq:henkm} by $g_i(z)$ and integrating:

\begin{equation}
    \tilde{\phi}^i_{avg}(q)=\tilde{\phi}^{i,0}_{avg}(q)/[1+q_s K_i(q)]
\end{equation}

where 

\begin{equation}
   K_i(q) = a_i\int_0^\infty dz \int_0^\infty dz'~ g_i(z) ~g(z')~ 2G_q(z,z') 
\end{equation}

Note that the product $a_i q_s$ defines an effectively layer-dependent Thomas-Fermi screening vector. 

It should be noted that the Thomas-Fermi approach has a clear limitation in the case of dichalcogenides. In these materials, both the charge density and the induced charge exhibit in-plane corrugations rather than being perfectly flat, because the Mo atoms occupy the trigonal prismatic sites between the chalcogen layers, effectively lying within the van der Waals gap. This leads to an overscreening of the disorder potential acting on the Mo-centered Wannier orbitals. For this reason, we applied a modification to Thomas-Fermi screening to obtain the potential on the Mo orbitals, described  in Sec. S8.

\section{Localization effects in a multi-valley BCS superconductor}
\label{app:c}

\subsection{Connection between the Wannier tight binding and momentum formulation in the presence of disorder in second quantization}

We rewrite the tight binding Hamiltonian in the presence of disorder in second quantization

\begin{equation}
H_{TB}
    = H_{sp} + H_{imp} = \sum_{\mathbf{R}_i\mathbf{R}_j} \sum_{l~m}
        H_{lm}(\mathbf{R}_i-\mathbf{R}_j) \,
        d_{\mathbf{R}_il}^{\dagger} d_{\mathbf{R}_jm}
    + \sum_{\mathbf{R}_il} U_{l}(\mathbf{R}_i) d_{\mathbf{R}_i,l}^{\dagger} d_{\mathbf{R}_il},
\end{equation}

where $H_{sp}$ is the clean Hamiltonian, $H_{imp}$ is the disorder Hamiltonian, \(H_{lm}(\mathbf{R}_i-\mathbf{R}_j)\) and \(U_l(\mathbf{R}_i)\) have been introduced above and \(d_{\mathbf{R_i}l}\) annihilates an electron in orbital \(l\) at lattice cell \(\mathbf{R}_i\). We Fourier-expand the creation and annihilation operators in the Bloch basis, $c_{\mathbf{k}l}$: 
\[
d_{\mathbf{R}_i l}^{\dagger}
    = \frac{1}{\sqrt{N}} \sum_{\mathbf{k}} e^{-i\mathbf{k}\cdot(\mathbf{R}_i+\boldsymbol{\tau}_l)} d_{\mathbf{k} l}^{\dagger},
\qquad
d_{\mathbf{R}_j m}
    = \frac{1}{\sqrt{N}} \sum_{\mathbf{k}'} e^{i\mathbf{k}'\cdot(\mathbf{R}_j+\boldsymbol{\tau}_m)} d_{\mathbf{k}' m},
\]

and substitute their expression in $H_{sp}$: 

\begin{equation}
H_{sp}
= \frac{1}{N} \sum_{l,m} \sum_{\mathbf{k},\mathbf{k}'} \sum_{\mathbf{R}_i,\mathbf{R}_j}
    H_{lm}(\mathbf{R}_i-\mathbf{R}_j)
    e^{-i\mathbf{k}\cdot(\mathbf{R}_i+\boldsymbol{\tau}_l)}
    e^{i\mathbf{k}'\cdot(\mathbf{R}_j+\boldsymbol{\tau}_m)} \,
    d_{\mathbf{k} l}^{\dagger} d_{\mathbf{k}' m}.
\end{equation}
We operate a change of variable from  $\mathbf{R}_i$ to \(\mathbf{R}=\mathbf{R}_i-\mathbf{R}_j\) and rewrite the phase factor as
\[
e^{-i\mathbf{k}\cdot(\mathbf{R}+\mathbf{R}_j+\boldsymbol{\tau}_l)}
e^{i\mathbf{k}'\cdot(\mathbf{R}_j+\boldsymbol{\tau}_m)}
= e^{-i\mathbf{k}\cdot\mathbf{R}} e^{-i(\mathbf{k}-\mathbf{k}')\cdot\mathbf{R}_j}
e^{-i\mathbf{k}\cdot\boldsymbol{\tau}_l} e^{i\mathbf{k}'\cdot\boldsymbol{\tau}_m}.
\]
By employing the lattice orthogonality, enforcing crystal-momentum conservation for the translationally-invariant hopping term, we obtain: 
\[
\sum_{\mathbf{R}_j} e^{-i(\mathbf{k}-\mathbf{k}')\cdot\mathbf{R}_j} = N \, \delta_{\mathbf{k},\mathbf{k}'},
\]

\begin{align}
H_{sp}
&= \sum_{\mathbf{k}} \sum_{l,m} \;  \left[
    \sum_{\mathbf{R}} H_{lm}(\mathbf{R}) \; e^{-i\mathbf{k}\cdot\mathbf{R}}
    \; e^{-i\mathbf{k}\cdot\boldsymbol{\tau}_l} e^{i\mathbf{k}\cdot\boldsymbol{\tau}_m}
    \right] d_{\mathbf{k} l}^{\dagger} d_{\mathbf{k} m} \nonumber \\
&= \sum_{\mathbf{k}} \sum_{l,m} \; \; H_{lm}(\mathbf{k}) \; d_{\mathbf{k} l}^{\dagger}  d_{\mathbf{k} m},
\label{eq:Hsp_k}
\end{align}

Fourier transforming $H_{imp}$ one obtains:
\begin{align*}
H_{imp}
&= \frac{1}{N}\sum_{l}\sum_{\mathbf{k},\mathbf{k}'} \sum_{\mathbf{R}_i}
    U_l(\mathbf{R}_i)\;
    e^{-i\mathbf{k}\cdot(\mathbf{R}_i+\boldsymbol{\tau}_l)}
    e^{+i\mathbf{k}'\cdot(\mathbf{R}_i+\boldsymbol{\tau}_l)}
    \; d_{\mathbf{k} l}^{\dagger} d_{\mathbf{k}' l} \\
&= \frac{1}{N}\sum_{l}\sum_{\mathbf{k},\mathbf{k}'} \left[
    \sum_{\mathbf{R}_i} U_l(\mathbf{R}_i)\, e^{i(\mathbf{k}'-\mathbf{k})\cdot(\mathbf{R}_i+\boldsymbol{\tau}_l)}
    \right] d_{\mathbf{k} l}^{\dagger} d_{\mathbf{k}' l}.
\end{align*}

We define the exchanged momentum $\mathbf{q}=\mathbf{k}-\mathbf{k}'$ and the disorder Fourier component

\begin{equation}\label{eq:Uq_def}
U_l(\mathbf{q})
    \equiv \sum_{\mathbf{R}} U_l(\mathbf{R}) \; e^{-i\mathbf{q}\cdot(\mathbf{R}+\boldsymbol{\tau}_l)}.
\end{equation}

and rewrite $H_{imp}$ as:
\begin{equation}
H_{imp}
    = \frac{1}{N}\sum_{l}\sum_{\mathbf{k},\mathbf{k}'}
        U_l(\mathbf{k}-\mathbf{k}') \; d_{\mathbf{k} l}^{\dagger} d_{\mathbf{k}' l}
    = \frac{1}{N}\sum_{l}\sum_{\mathbf{k},\mathbf{q}}
        U_l(\mathbf{q}) \; d_{\mathbf{k}+\mathbf{q}l}^{\dagger} d_{\mathbf{k} l},
\label{eq:disorder}
\end{equation}
where \(\mathbf{q}=\mathbf{k}-\mathbf{k}'\) is the momentum transfer. The tight binding Hamiltonian is thus written in the form: 

\begin{equation}\label{eq:tb}
H_{TB} = \sum_{\mathbf{k}}\sum_{l,m} H_{lm}(\mathbf{k})  d_{\mathbf{k} l}^{\dagger} d_{\mathbf{k} m} + \frac{1}{N} \sum_{l} \sum_{\mathbf{k}\mathbf{k}'} U_l(\mathbf{k}-\mathbf{k}') d_{\mathbf{k}' l}^{\dagger} d_{\mathbf{k} l} 
\end{equation}

We now operate a unitary rotation from the Wannier to the Bloch basis:

\begin{equation}\label{eq:evecs}
\sum_{m} H_{lm}(\mathbf{k}) \, u_{m n}(\mathbf{k}) = \varepsilon_{n}(\mathbf{k}) \, u_{l n}(\mathbf{k}),
\end{equation}

and define band annihilation operators \(c_{\mathbf{k} n}\) by projecting onto the Bloch eigenvectors:
\begin{equation}\label{eq:band_op}
c_{\mathbf{k} n} \;=\; \sum_{l} u_{l n}^{*}(\mathbf{k}) \; d_{\mathbf{k} l}.
\end{equation}

$H_{sp}$ becomes: 

\begin{align}
H_{sp}= \sum_{l,m} H_{lm}(\mathbf{k}) \, d_{\mathbf{k} l}^{\dagger} d_{\mathbf{k} m}
&= \sum_{l,m} \sum_{n,n'} H_{lm}(\mathbf{k}) \, u_{l n}(\mathbf{k}) \, u_{m n'}^{*}(\mathbf{k}) \;
    c_{\mathbf{k} n}^{\dagger} c_{\mathbf{k} n'} \nonumber\\
&= \sum_{n} \varepsilon_{\mathbf{k}n} \; c_{\mathbf{k} n}^{\dagger} c_{\mathbf{k} n}.
\end{align}

The disorder term instead becomes: 
\begin{align}
H_{imp}
&= \frac{1}{N}\sum_{l}\sum_{\mathbf{k},\mathbf{k}'}
    U_l(\mathbf{k}-\mathbf{k}') \; d_{\mathbf{k}',l}^{\dagger} d_{\mathbf{k} l} \nonumber\\
&= \frac{1}{N}\sum_{\mathbf{k},\mathbf{k}'} \sum_{l} U_l(\mathbf{k}-\mathbf{k}')
    \left( \sum_{n'} u_{l n'}^{*}(\mathbf{k}') \, c_{\mathbf{k}',n'}^{\dagger} \right)
    \left( \sum_{n} u_{l n}(\mathbf{k}) \, c_{\mathbf{k} n} \right) \nonumber\\
&= \frac{1}{N}\sum_{\mathbf{k},\mathbf{k'}} \sum_{n,n'} 
    \sum_{l} U_l(\mathbf{k}-\mathbf{k}') \, u_{l n}^{*}(\mathbf{k}') \, u_{l n'}(\mathbf{k})  
    \; c_{\mathbf{k}',n'}^{\dagger} c_{\mathbf{k} n}.
\label{eq:Himp_band}
\end{align}

Define the band-basis disorder matrix elements
\begin{equation}\label{eq:Vndef}
V_{n'\mathbf{k}'  n\mathbf{k}}
    \equiv \dfrac{1}{N}\sum_{l} U_l(\mathbf{k}-\mathbf{k}') \; u_{l n'}^{*}(\mathbf{k}') \; u_{l n}(\mathbf{k}).
\end{equation}

Thus the disordered Wannier tight-binding Hamiltonian in the band basis reads

\begin{equation}\label{eq:H_full_band}
H_{TB}
    = \sum_{\mathbf{k}}\sum_{l} \varepsilon_{\mathbf{k}l} \; c_{\mathbf{k} l}^{\dagger} c_{\mathbf{k} l}
    \;+\; \sum_{\mathbf{k},\mathbf{k}'} \sum_{l,l'}
        V_{l'\mathbf{k}' l\mathbf{k}} \;
        c_{\mathbf{k'},l'}^{\dagger} c_{\mathbf{k} l}.
\end{equation}

which is the starting point of out derivation of the linearized multi-gap equation below.

\subsection{Derivation of the linearized multi-gap equation in linear response}
In this Appendix~we consider the effect of localization on the critical temperature and superconducting gap of a multi-valley BCS superconductor. Our discussion builds on top of the results by Maekawa and Fukuyama for a BCS superconductor in the presence of disorder\cite{doi:10.1143/JPSJ.51.1380}. In our analysis we will neglect spin-orbit coupling (SOC) effects on the disorder corrections. Since SOC is sizable in the materials discussed here its effects may be important and neglecting them may not be justified. Aware of this approximation we continue our analysis and leave the discussion of SOC as a possible future extension of this work.  We consider the following BCS mean field Hamiltonian in real space:

%The multiband picture discussed in the following is physically justified if one assumes that disorder does not mix states among bands, whereas the single-band treatment of Ref.\cite{doi:10.1143/JPSJ.51.1380} is sufficient to describe the superconducting state if disorder mixes all bands and no meaningful band quantum number can be defined.
%%ATTENZIONE METTERE -ALPHA CIOE BLOCH E METTERE ESPRESSIONE ESPLICITA DEL KN INVECE DI ALPHA CHE E' + OSCURO
\begin{equation}
    H = H_0 + H_{BCS} = H_0 -  \ ~ \int d \mathbf{r}' \int d\mathbf{r} ~ V^{BCS}(\mathbf{r},\mathbf{r}')[<\Psi_{\downarrow}^\dagger(\mathbf{r'})\Psi_{\uparrow}^\dagger(\mathbf{r'})>_H \Psi_{\uparrow}(\mathbf{r})\Psi_{\downarrow}(\mathbf{r})+ <\Psi_{\uparrow}(\mathbf{r'})\Psi_{\downarrow}(\mathbf{r'})>_H\Psi_{\downarrow}^\dagger(\mathbf{r})\Psi_{\uparrow}^\dagger(\mathbf{r})]
    \label{eq:1sc}
\end{equation}

having defined the electronic field operators $\Psi_{\sigma}$  with spin $\sigma$, and the effective superconducting potential $V^{BCS}(\mathbf{r},\mathbf{r}')$. We also defined the thermal average:

\begin{equation}
  <\Psi_{\uparrow}(\mathbf{r})\Psi_{\downarrow}(\mathbf{r})>_H = \int d\mathbf{r} \dfrac{~\mathrm{Tr}~ e^{-\beta H} \Psi_{\uparrow}(\mathbf{r})\Psi_{\downarrow}(\mathbf{r})}{\mathrm{Tr}~ e^{-\beta H}}   
\end{equation}

and the Hamiltonian H$_0$:

\begin{equation}
    H_0 = H_{sp}+H_{imp}+H_{el-el}
\end{equation}

where $H_{sp}$ and $H_{imp}$ are the already discussed Bloch Hamiltonian and disordered potential. $H_{el-el}$ is the two-body electron-electron Hamiltonian:

\begin{equation}
\begin{gathered}
\hat{H}_{el-el} = \dfrac{1}{2N} \sum_{\mathclap{\substack{\vb{k,k',q} \\ l,l',m,m' }}} W_{\vb{kk'k-qk'+q}}^{ll'mm'} \hat{c}^\dagger_{\vb{k}l}\hat{c}^\dagger_{\vb{k'}l'}\hat{c}_{\vb{k'+q}m'}\hat{c}_{\vb{k-q}m}, 
\end{gathered}
\end{equation}
where the matrix element of the screened Coulomb potential is defined as
\begin{equation}
    W_{\vb{kk'k-qk'+q}}^{ll'mm'} = \mel{\psi_{\vb{k}l}\psi_{\vb{k'}l'}}{w(|\hat{\vb{r}}-\hat{\vb{r}}'|)}{{\psi_{\vb{k-q}m}\psi_{\vb{k'+q}m'}}}.
\end{equation}

and $w(|\hat{\vb{r}}-\hat{\vb{r}}'|)$ is the screened Coulomb potential. 
We assume that the single-particle disorder Hamiltonian $H_{imp}$, which is not diagonal in $\mathbf{k}$, remains block-diagonal in certain composite Bloch sectors  $\gamma$,  which spans a subset of $\{ \mathbf{k}, l \}$ states. In the case of gated dichalcogenides, the index $\gamma$ labels valleys, having in mind that each valley can be composed of several bands (e.g. bands having the same minimum/maximum, represented by the index $l$). These assumptions amount to neglecting inter-valley matrix elements ($\gamma \neq \gamma'$) in Eq.\ref{eq:Vndef} and only retaining intraband matrix elements inside the same valley. The disorder  Hamiltonian is then restricted to the block-diagonal form:
\begin{equation}
    H_{imp} = 
    \sum_{\gamma}
    \sum_{\mathbf{k},l \in \gamma}
    \sum_{\mathbf{k}',l' \in \gamma}
    V^{\gamma}_{\mathbf{k}l,\mathbf{k}'l'}\,
    c_{\mathbf{k}l}^\dagger c_{\mathbf{k}'l'} ,
\end{equation}

where $V^{\gamma}_{\mathbf{k}n,\mathbf{k}'n'}$ represents the disorder matrix elements within block $\gamma$. This assumption is justified when $V^{\gamma}_{\mathbf{k}n,\mathbf{k}'n'}$, Eq.\ref{eq:Vndef}, decays quickly with $\mathbf{q}=\mathbf{k}-\mathbf{k}'$. Without loss of generality, we expand the electron field operator in the complete basis set that diagonalizes the single-particle Hamiltonian $H_{sp}+H_{imp}$. We shall assume that states in this set can be labeled by two quantum numbers, a first quantum number $n$ and a second valley quantum number $\gamma$.  %and is well motivated at low doping with small Thomas-Fermi wave vector, less so at high doping, and it is worth considering since electrons in $Q$ valleys have larger projections on Sulfur $p$ orbitals than electrons in $K$ valleys (see Fig.~\ref{figS1}), thus spending more time closer to the gate-induced disordered potential. 

\begin{equation}
   \Psi_{\sigma}(\mathbf{r}) =   \sum_{\gamma=1}^{N_v} \sum_{n=1}^{N_\gamma}\phi_{n}^{\gamma}(\mathbf{r}) c_{n\sigma}^{\gamma} \end{equation}

   where the annihilation operator $c^\gamma_{n \sigma}$ destroys an electron with wavefunction $\phi^\gamma_n(\mathbf{r})$, $N_v$ is the number of valleys and $N_\gamma$ is the number of states in the band $\gamma$. $H_{BCS}$ becomes

\begin{equation}
\begin{gathered}
 H_{BCS} =  -  \ ~ \int d \mathbf{r}' \int d\mathbf{r} ~ V^{BCS}(\mathbf{r},\mathbf{r}')[\sum_{nn'\gamma\gamma'}\phi^{\gamma*}_n(\mathbf{r'})\phi^{\gamma'*}_{n'}(\mathbf{r'})<c_{n\downarrow}^{\gamma\dagger} c_{n'\uparrow}^{\gamma'\dagger} >_H \sum_{n''n'''\gamma{''} \gamma{'''}}\phi_{n''}^{\gamma''}(\mathbf{r})\phi^{\gamma'''}_{n'''}(\mathbf{r})c_{n''\uparrow}^{\gamma''} c^{\gamma'''}_{n'''\downarrow} + \\\sum_{nn'\gamma\gamma'}\phi^{\gamma}_n(\mathbf{r'})\phi^{\gamma'}_{n'}(\mathbf{r'})<c_{n\uparrow}^{\gamma~} c_{n'\downarrow}^{\gamma'~} >_H \sum_{n''n'''\gamma{''} \gamma{'''}}\phi_{n''}^{\gamma''*}(\mathbf{r})\phi^{\gamma'''*}_{n'''}(\mathbf{r})c_{n''\downarrow}^{\gamma''\dagger} c^{\gamma'''\dagger}_{n'''\uparrow} ]
    \label{eq:2sc}
    \end{gathered}
\end{equation}

where we left the summations' range implicit. ~We define the BCS potential matrix element as:

\begin{equation}
   V^{BCS}_{n n' n'' n'''\gamma \gamma' \gamma '' \gamma'''} =\ ~ \int d \mathbf{r}' \int d\mathbf{r} ~\phi_n^{\gamma*}(\mathbf{r'})\phi_{n'}^{\gamma'*}(\mathbf{r'})~V^{BCS}(\mathbf{r},\mathbf{r'})~\phi^{\gamma''}_{n''}(\mathbf{r})\phi^{\gamma'''}_{n'''}(\mathbf{r})
\end{equation}

which up to now couples both states in the same valley and states in different valleys. We shall now assume a diagonal pairing:

\begin{equation}
    V^{BCS}_{n n' n '' n ''' \gamma \gamma' \gamma '' \gamma'''} = V^{BCS}_{n n''\gamma \gamma''}\delta_{\gamma\gamma'}\delta_{\gamma''\gamma'''}\delta_{n n'^{*}}\delta_{n'' n'''^{*}}=V^{BCS}_{\gamma\gamma'nn'}
\end{equation}

where  we are labeling with $n^*$ the time-reversal conjugate of state $n$.

Eq. \ref{eq:2sc} becomes:

\begin{equation}
     H_{BCS} =  - \sum_{n\gamma}\sum_{n'\gamma'} V^{BCS}_{\gamma\gamma'nn'} <c_{n^*\downarrow}^{\gamma\dagger} c_{n\uparrow}^{\gamma\dagger} >_H  c_{n'\uparrow}^{\gamma'} c^{\gamma'}_{n'^*\downarrow} - \sum_{n\gamma}\sum_{n'\gamma'}  V^{*BCS}_{\gamma\gamma'nn'} <c_{n\uparrow}^{\gamma} c_{n^*\downarrow}^{\gamma} >_H  c_{n'^*\downarrow}^{\gamma'\dagger} c^{\gamma'\dagger}_{n'\uparrow}
\end{equation}

 We are interested in the gap function:

\begin{equation}
    \Delta_{n\gamma}=\sum_{n'\gamma'} V^{BCS}_{n n'\gamma \gamma'}<c_{n'\uparrow}^{\gamma'} c_{n'^*\downarrow}^{\gamma'}>_H \label{eq:BCS}\end{equation}

The evaluation of the correlator in the right hand side of Eq.~\ref{eq:BCS},$<c_{n'}^{\gamma'} c_{n'}^{\gamma'}>_H$ , is performed in imaginary time linear response, treating $H_{BCS}$ as the perturbation. We find:

\begin{equation}
\begin{gathered}
   <c_{n'\uparrow}^{\gamma'} c_{n'\downarrow}^{\gamma'}>_H = - \sum_{\gamma\gamma''}\sum_{nn''}V^{*BCS}_{nn''\gamma\gamma''}<c^\gamma_{n\uparrow}c^\gamma_{n^*\downarrow}>_H\int_0^\beta d\tau' <T_O  c^{\gamma'}_{ n'\uparrow}(0)c^{\gamma'}_{ n'^*\downarrow}(0)c_{ n''^*\downarrow}^{\gamma''\dagger}(\tau')c_{ n''\uparrow}^{\gamma''\dagger}(\tau')>_{H_0} 
    \label{eq:delta}
    \end{gathered}
\end{equation}

where the thermal average is intended on $H_0$, $T_O$ denotes the time ordering operator and $\beta$ is the inverse temperature.

%\begin{equation}
%\begin{gathered}
%   <c_{n'*\uparrow}^{\gamma'} c_{n'\downarrow}^{\gamma'}>_H = - \sum_{\gamma}\sum_{n}V^{BCS}_{n'n\gamma'\gamma}<c^\gamma_{n\uparrow}c^\gamma_{n\downarrow}>_H\int_0^\beta d\tau' <T_O  c^{\gamma'}_{ n'\uparrow}(0)c^{\gamma'}_{ n'\downarrow}(0)c_{ n'\downarrow}^{\gamma'\dagger}(\tau')c_{ n'\uparrow}^{\gamma'\dagger}(\tau')>_{H_0} 
%    \label{eq:delta2}
%    \end{gathered}
%\end{equation}

where we used that $V^{*BCS}_{nn'\gamma\gamma'}=V^{BCS}_{n'n\gamma'\gamma}$. We recognize in Eq.\ref{eq:delta} the superconducting gap definition and rewrite it as:

\begin{equation}
\begin{gathered}
   <c_{n'\uparrow}^{\gamma'} c_{n'^*\downarrow}^{\gamma'}>_H = - \Delta_{n'\gamma'} \sum_{n'' \gamma''}\int_0^\beta d\tau' <T_O  c^{\gamma'}_{ n'\uparrow}(0)c^{\gamma'}_{ n'^*\downarrow}(0)c_{ n''^*\downarrow}^{\gamma''\dagger}(\tau')c_{ n''\uparrow}^{\gamma''\dagger}(\tau')>_{H_0} 
    \label{eq:delta3}
    \end{gathered}
\end{equation}

and replacing the expression for $ <c_{n'\uparrow}^{\gamma'} c_{n'\downarrow}^{\gamma'}>_H$  in the definition of $\Delta_{n~\gamma}$, Eq.\ref{eq:delta}, we find the self consistent expression:

\begin{equation}
      \Delta_{n\gamma}=-\sum_{n'\gamma'} V^{BCS}_{n n'\gamma \gamma'} \Delta_{n'\gamma'}\sum_{n''\gamma''}\int_0^\beta d\tau' <T_O  c^{\gamma'}_{ n'\uparrow}(0)c^{\gamma'}_{ n'^*\downarrow}(0)c_{ n''^*\downarrow}^{\gamma''\dagger}(\tau')c_{ n''\uparrow}^{\gamma''\dagger}(\tau')>_{H_0} \label{eq:BCSf}
\end{equation}

We shall assume intraband isotropy, $i.e.$ 

\begin{equation}
\begin{gathered}
        V^{BCS}_{n n'\gamma \gamma'} = V^{BCS}_{\gamma \gamma'}\\
        \Delta_{n~\gamma} = \Delta_\gamma
\end{gathered}
\end{equation}

The multi-band gap equation becomes:
\begin{equation}
\begin{gathered}
  \Delta_\gamma =  - \sum_{\alpha}V^{BCS} _{\gamma\alpha}\Delta_{\alpha}\sum_{\alpha'nn'}\int_0^\beta d\tau' <T_O  c_{\alpha n\uparrow}^{\alpha}(0)c_{n^*\downarrow}^{\alpha}(0)c_{ n'^*\downarrow}^{\alpha' \dagger}(\tau')c_{n'\uparrow}^{\alpha'\dagger}(\tau')>_{H_0} \\\\
    \label{eq:delta2}
    \end{gathered}
\end{equation}

%We shall generally assume that the only non-vanshing terms are intra-band $<T_O  c^{\gamma'}_{ n'\uparrow}(0)c^{\gamma'}_{ n'\downarrow}(0)c_{ n'\downarrow}^{\gamma'\dagger}(\tau')c_{ n'\uparrow}^{\gamma'\dagger}(\tau')>_{H_0}$  are non-vanishing in Eq.\ref{eq:delta}. This approximation is physically motivated if one assumes that intraband scattering is generally much stronger 
In the absence of electron-electron interactions ($i.e.$ if $H_{el-el}=0$) the time average can be factorized as:

\begin{equation}
     <T_O  c_{n\uparrow}^{\alpha}(0)c_{n^*\downarrow}^{\alpha}(0)c_{n'^*\downarrow}^{\alpha'\dagger}(\tau')c_{n'\uparrow}^{\alpha'\dagger}(\tau')>_{H_0}=\delta_{\alpha \alpha'}\delta_{nn'} <T_Oc_{n^*\downarrow}^{\alpha}(0)c_{ n'^*\downarrow}^{\alpha'\dagger}(\tau')>_{H_0}<T_Oc_{ n\uparrow}^{\alpha}(0)c_{n'\uparrow}^{\alpha'\dagger}(\tau')>_{H_0}
\end{equation}
and one is left with the following expression:

\begin{equation}
      \Delta_\gamma =  -\sum_{\alpha}V^{BCS}_{\gamma\alpha} \Delta_{\alpha} \sum_{n=1}^{N_{\alpha}}\int_0^\beta d\tau' <T_Oc_{n^*\downarrow}^\alpha(0)c_{n^*\downarrow}^{\alpha\dagger}(\tau')>_{H_0}<T_Oc_{n\uparrow}^\alpha(0)c_{n\uparrow}^{\alpha\dagger}(\tau')>_{H_0} \\\\
\end{equation}

Introducing the imaginary time Green's function in the Matsubara frequency representation one has:
\begin{equation}
\begin{gathered}
    \Delta_\gamma = -\sum_{\alpha } V^{BCS}_{\gamma\alpha}\Delta_{\alpha}\sum_{n=1}^{N_{\alpha}} G_{\alpha n}(\omega_m)G_{\alpha n}(-\omega_{m}) \\
   \Rightarrow \Delta_\gamma= \sum_{\alpha} V^{BCS}_{\gamma\alpha}\Delta_{\alpha}\sum_{n=1}^{N_{\alpha}} \sum_{m=-\infty}^{\infty} \dfrac{1}{\omega_m^2+\varepsilon_n^2}
    \end{gathered}
    \label{eq:linearBCS}
\end{equation}

where we introduced the Matsubara frequency $\omega_m = 2\pi T(n+\frac12)$ . This is the spatially isotropic linearized BCS gap equation for the multi-valley case in the presence of impurities. A more familiar form is recovered employing the identity (cfr Chapter 2 of Ref.\cite{ALLEN19831}):

\begin{equation}
    \sum_{m=-\infty}^\infty [(2m+1)^2\pi^2+a^2]^{-1}=2a^{-1}\tanh(\frac{a}{2})
\end{equation}

From this results it is clear that the Anderson's theorem still holds in a multi-valley superconductor as long as the impurity Hamiltonian doesn't mix different bands.

When strong disorder is present and $H_{el-el}$ is not neglected, the pair propagator in the right hand side of Eq.~\ref{eq:delta2} acquires additional contributions of the order of $(\varepsilon_F\tau_0)^{-1}$ due to the interplay between disorder and Coulomb interaction, causing a renormalization of the superconducting critical temperature. This aspect was studied within perturbation theory in Ref.\cite{doi:10.1143/JPSJ.51.1380}, where it was shown that logarithmic corrections to the critical temperature $T_c$ appear, leading to a reduced $T_{c}$. In the next two subsections we discuss the form of this correction both for the single valley and the multi-valley case.

\subsection{Disorder renormalization of $T_c$ in the single valley case}

In the presence of strong interband disorder, the band anisotropy is washed out and we shall assume that the superconducting state is well described by a single valley equation. This scenario is recovered from Eq.~\ref{eq:linearBCS} if $V_{\gamma \alpha}^{BCS}=V^{BCS}$ and thus $\Delta_{\gamma}=\Delta$. We obtain the usual linearized BCS gap equation:

\begin{equation}
    \Delta = \Delta \sum_{n=1}^{N}\dfrac{V^{BCS}}{2 \varepsilon_n}\tanh(\dfrac{\varepsilon_n}{2T})
\end{equation}

This linear equation always possesses the solution $\Delta=0$. A non trivial $\Delta \neq 0$ solution becomes possible at $T=T_c$\cite{ALLEN19831}:

\begin{equation}
    T_c = 1.13~\theta_D \exp[-1/N(0)V^{BCS}]
\end{equation}

where $\theta_D$ is an energy cutoff given by the range of $V^{BCS}$ for the $n$ summation and $N(0)$ is the density of states at Fermi level. This result is perfectly in the spirit of the Anderson's theorem\cite{ANDERSON195926,doi:10.1143/JPSJ.51.1380}. If we perturbatively consider the effect of the electron-electron interaction in the presence of disorder to the lowest order of $(\varepsilon_F\tau_0)^{-1}$, the pair propagator in the right hand side of Eq.~\ref{eq:delta2} acquires additional contributions due to impurity electron-hole and electron-electron vertex\cite{doi:10.1143/JPSJ.51.1380}. These corrections to the pair propagator can be schematically separated in two groups: the Hartree-Fock corrections that cause square logarithmic corrections $\propto\ln(1/T)^2$ and vertex corrections that lead to a cubic logarithmic corrections $\propto\ln(1/T)^3$, which represents the dominant correction term to the lowest order.
In the single valley case, Eq.$\ref{eq:delta2}$ becomes:

\begin{equation}
     \Delta= \Delta~V^{BCS}[ \sum_{n=1}^{N} \sum_{m=-\infty}^{\infty} \dfrac{1}{\omega_m^2+\varepsilon_n^2}+R_{HF}(T)  +R_{v}(T)]
\end{equation}

where we took $V^{BCS}$ out of the $n$ summation with the implicit understanding that the $n$ summation only extends in the energy range where $V^{BCS}$ is nonzero. These corrections have been evaluated for the two-dimensional case in Ref.\cite{doi:10.1143/JPSJ.51.1380}, obtaining:

\begin{equation}
R_{HF}(T)=-\dfrac{(g_1-3g')N(0)^2}{4\pi\varepsilon_F \tau_0}(\ln \dfrac{1}{T \tau_0})^2    
\end{equation}

and 

\begin{equation}
    R_v(T) = -\dfrac{(g_1+g')N(0)^2}{6\pi\varepsilon_F\tau_0}(\ln\dfrac{1}{T\tau_0})^3
\end{equation}

here $\tau_0$ is the scattering time, $\varepsilon_F$ is the Fermi energy, $g_1$ is a constant that depends on the electron-electron processes ($g_1 N(0)=1/2$ for screened Coulomb interaction in the 2D homogeneous electron gas\cite{FINKELSTEIN1994636}) and $g'$ is a constant related to electron-phonon processes, which we consider negligible with respect to the electron-electron ones\cite{doi:10.1143/JPSJ.51.1380}. Note that a factor $N(0)$ is missing in the definitions given in Eqs.(26) and (27) of Ref.\cite{doi:10.1143/JPSJ.51.1380}.This treatment is only valid in the perturbative regime\cite{FINKELSTEIN1994636} and overestimates the $T_c$ renormalization when $\delta T_c \approx T_{c0}$. In this case, a renormalization group approach must be used to study the (possible) re-entrant Anderson insulating phase. 

In this case, the usual linearized BCS treatment gives a self-consistent equation for the critical temperature $T_c$\cite{doi:10.1143/JPSJ.51.1380,ALLEN19831}:

\begin{equation}
\begin{gathered}
        \Delta = \Delta ~V^{BCS}[T \sum_{m=-\infty}^{\infty} \pi N(0)/|\omega_m|+R_{HF}(T_c)+R_{v}(T_c)]\\
        \Rightarrow \dfrac{1}{V^{BCS}} = [\ln(\theta_D/2 \pi T_c)-\ln(e^\gamma/4)]N(0)+R_{HF}(T_c)+R_v(T_c)
\end{gathered}
\end{equation}

After some straightforward manipulations one has:

\begin{equation}
    \dfrac{1}{V^{BCS}N(0)} \approx [ 0.1254+~\ln(\theta_D/ T_c)-\dfrac{(g_1-3g')N(0)}{4\pi\varepsilon_F \tau_0}(\ln \dfrac{1}{T_c \tau_0})^2-\dfrac{(g_1+g')N(0)}{6\pi\varepsilon_F\tau_0}(\ln\dfrac{1}{T_c\tau_0})^3]\label{eq:Tc}
\end{equation}

By subtracting this equation to the equation for the original $T_{c0}$ one obtains:

\begin{equation}
    \ln(T_c/T_{c0}) = [ -\dfrac{(g_1-3g')N(0)}{4\pi\varepsilon_F \tau_0}(\ln \dfrac{1}{T_c \tau_0})^2-\dfrac{(g_1+g')N(0)}{6\pi\varepsilon_F\tau_0}(\ln\dfrac{1}{T_c\tau_0})^3] \label{eq:Tciso}
\end{equation}

$i.e.$ a self-consistent relationship for $T_c$ suppression\cite{doi:10.1143/JPSJ.51.1380,FINKELSTEIN1994636,1987PZETF4537F,Larkin2008}, which gives the critical temperature renormalization for a two dimensional single gap BCS superconductor in the presence of disorder in the perturbative regime.

\subsection{Disorder renormalization of $T_c$ in the multi-valley case}

We now consider Eq.~\ref{eq:linearBCS}  in the multi-valley case. Proceeding in a very similar same way we obtain a system of equations:

\begin{equation}
            \Delta_{\gamma} = \sum_{\alpha}\Delta_{\alpha}V^{BCS}_{\gamma \alpha} [T \sum_{m=-\infty}^{\infty} \pi N_{\alpha}(0)/|\omega_m|+R^{\alpha}_{HF}(T)+R^{\alpha}_{v}(T)]
            \label{eq:matrixBCS}
\end{equation}

where we introduced the multi-gap renormalization factors

\begin{equation}
\begin{gathered}
R_{HF}^\alpha(T)=-\dfrac{(g_1-3g')N_\alpha(0)^2}{4\pi\varepsilon^\alpha_F \tau^\alpha_0}(\ln \dfrac{1}{T \tau^\alpha_0})^2    \\
    R_v^\alpha(T) = -\dfrac{(g_1+g')N_\alpha(0)^2}{6\pi\varepsilon^\alpha_F\tau^\alpha_0}(\ln\dfrac{1}{T\tau^\alpha_0})^3
    \end{gathered}
\end{equation}

where $N_\alpha(0)$ is the $\alpha$-band restricted density of states, $\tau_0^\alpha$ is the scattering time for the $\alpha$ band and $\varepsilon_F^\alpha$ is the Fermi energy of the $\alpha$ band. Again, we are assuming that disorder plus interaction corrections only give intraband scattering, a picture that could be physically meaningful if the intraband Coulomb interaction is much stronger than the interband one. This is physically justified by previous studies on the Coulomb interaction in doped dichalcogenides\cite{PhysRevB.94.134504}. The system of equations Eq.~\ref{eq:matrixBCS} has the form

\begin{equation}
    \Delta_\gamma = \sum_{\alpha}K_{\gamma \alpha}(T)\Delta_\alpha
\end{equation}

where $K_{\alpha \gamma}$ is the renormalized BCS kernel:
\begin{equation}
    K_{ \gamma \alpha}(T) = V^{BCS}_{\gamma \alpha} [T \sum_{m=-\infty}^{\infty} \pi N_{\alpha}(0)/|\omega_m|+R^{\alpha}_{HF}(T)+R^{\alpha}_{v}(T)]
\end{equation}

 In the perturbative regime $K_{\gamma,\alpha}$>0 and $\Delta$ is a positive eigenvector. In this case, the Perron-Frobenius theorem\cite{YANG200560} allow us to to write the equation for $T_c$ as:

\begin{equation}
    \kappa^{max}(T_c) = 1
\end{equation}

where $\kappa^{max}(T_c)$ represents the maximum eigenvalue of the kernel $K(T=T_c)$. 

W discuss in some more detail the effects of the disorder corrections in the multi-gap case. First, we perform the Matsubara summation and arrive to the following expression for the kernel:

\begin{equation}
    K_{\gamma \alpha} = V^{BCS}_{\gamma \alpha}N_\alpha(0)[0.1254 +\ln(\theta_D/T_c)-\dfrac{(g_1-3g')N_\alpha(0)}{4\pi\varepsilon^\alpha_F \tau^\alpha_0}(\ln \dfrac{1}{T_c\tau^\alpha_0})^2 -\dfrac{(g_1+g')N_\alpha(0)}{6\pi\varepsilon^\alpha_F\tau^\alpha_0}(\ln\dfrac{1}{T_c\tau^\alpha_0})^3]
\end{equation}

The disorder induced corrections do not change the structure of the kernel. We focus on the two band case for definiteness, where we can find more insight on the supercondcuting gap of the disordered system and write an analytic expression for the $T_c$ correction.

\begin{equation}
\begin{gathered}
 \Delta_{1}=\lambda_{11}\Delta_1[\ln(\theta_D/T_c)+\alpha_1(T_c)]+\lambda_{12}\Delta_2[\ln(\theta_D/T_c)+\alpha_2(T_c)]   \\
  \Delta_{2}=\lambda_{22}\Delta_2[\ln(\theta_D/T_c)+\alpha_2(T_c)]+\lambda_{21}\Delta_1[\ln(\theta_D/T_c)+\alpha_1(T_c)] 
  \label{eq:BCS2b}
\end{gathered}
\end{equation}

where we introduced the function
\begin{equation}
    \alpha_{1,2}(T_c)=0.1254-\dfrac{(g_1-3g')N_{1,2}(0)}{4\pi\varepsilon^{1,2}_F\tau_0^{1,2}}(\ln \dfrac{1}{T_c\tau_0^{1,2}})^2 -\dfrac{(g_1+g')N_{1,2}(0)}{6\pi\varepsilon^{1,2}_F\tau_0^{1,2}}(\ln\dfrac{1}{T_c\tau^{1,2}_0})^3
\end{equation}

and the electron-phonon coupling parameter $\lambda_{\gamma,\alpha} = V^{BCS}_{\gamma,\alpha}N(0)_\alpha$
\begin{equation}
\begin{gathered}
 \ln(\theta_D/T_c)(\lambda_{11}\Delta_1+\lambda_{12}\Delta_2)=(1-\lambda_{11}\alpha_1(T_{c}))\Delta_1-\alpha_2(T_c)\lambda_{12}\Delta_2\\
  \ln(\theta_D/T_c)(\lambda_{22}\Delta_2+\lambda_{21}\Delta_1)=(1-\lambda_{22}\alpha_2(T_{c}))\Delta_2-\alpha_1(T_c)\lambda_{21}\Delta_1
\end{gathered}
\end{equation}

that gives the equation

\begin{equation}
    \dfrac{(1-\lambda_{11}\alpha_1(T_{c}))\Delta_1-\alpha_2(T_c)\lambda_{12}\Delta_2}{(\lambda_{11}\Delta_1+\lambda_{12}\Delta_2)}=\dfrac{(1-\lambda_{22}\alpha_2(T_{c}))\Delta_2-\alpha_1(T_c)\lambda_{21}\Delta_1}{(\lambda_{22}\Delta_2+\lambda_{21}\Delta_1)}
\end{equation}

This gives a quadratic equation of the form:

\begin{equation}
    \lambda_{12}\Delta_2^2+b(T_c)\Delta_1\Delta_2-\lambda_{21}\Delta_1^2=0
\end{equation}

having defined:

\begin{equation}
        b(T_c) = \lambda_{11}-\lambda_{22}+{\lambda_{11}\lambda_{22}}(\alpha_1(T_c)-\alpha_2(T_c))\\
\end{equation}

We choose the solution that gives the same sign for $\Delta_1$ and $\Delta_2$:

\begin{equation}
    \Delta_2 = \dfrac{-b(T_c)+\sqrt{b(T_c)^2+4\lambda_{12}\lambda_{21}}}{2\lambda_{12}}~\Delta_1 = F(T_c)~\Delta_1
    \label{eq:gaprelation}
\end{equation}

The relationship in Eq.~\ref{eq:gaprelation} shows that anisotropic band disorder ($\alpha_1 \neq \alpha_2$) can induce multi-gap features even in an otherwise isotropic superconductor. 

Substituting this expression for $\Delta_2$ in the first line of Eq.~\ref{eq:BCS2b} one obtains:

\begin{equation}
    1 = (\lambda_{11}+\lambda_{12}F(T_c))\ln(\dfrac{\theta_D}{T_c})+\lambda_{11}\alpha_1(T_c)+\lambda_{12}F(T_c)\alpha_2(T_c)
\end{equation}

and finally come to an expression for $T_c$ in a two band superconductor in the presence of disorder:

\begin{equation}
\begin{gathered}
       \ln(T_c/\theta_D)= -\dfrac{1-\lambda_{11}\alpha_1(T_c)+\lambda_{12}F(T_c)\alpha_2(T_c)}{\lambda_{11}+\lambda_{12}F(T_c)}
\end{gathered}
\end{equation}

We can then formulate a self-consistent equation for $T_c$ with the same structure of the one discussed for the single valley case, Eq.~\ref{eq:Tciso}\cite{doi:10.1143/JPSJ.51.1380}:

\begin{equation}
\begin{gathered}
       \ln(T_c/T_{c0})= -\dfrac{1-\lambda_{11}\alpha_1(T_c)+\lambda_{12}F(T_c)\alpha_2(T_c)}{\lambda_{11}+\lambda_{12}F(T_c)}+\dfrac{1-\lambda_{11}\alpha_1(T_{c0})-\lambda_{12}F(T_{c0})\alpha_2(T_{c0})}{\lambda_{11}+\lambda_{12}F(T_{c0})}
       \label{eq:Tcaniso}
\end{gathered}
\end{equation}

It is straightforward to verify that Eq.~\ref{eq:Tcaniso} reduces to Eq.~\ref{eq:Tciso} when $F(T_{c0})=1$. The maximum anisotropy is reached when $F(T_c)\rightarrow~0$ or $F(T_c)\rightarrow~\infty$, when one has:

\begin{equation}
\begin{gathered}
     \ln(T_c/T_{c0}) =\alpha_1(T_c)-\alpha_1(T_{c0})\hspace{2cm} F(T_c)\rightarrow~0\\
          \ln(T_c/T_{c0}) = \alpha_2(T_c)-\alpha_2(T_{c0}) \hspace{2cm} F(T_c)\rightarrow~\infty
\end{gathered}
\end{equation}

reflecting the fact that when one band dominates superconductivity it completely determines $T_c$ also in the presence of the corrections.

\begin{figure}[t]
\includegraphics[width=0.7\columnwidth]{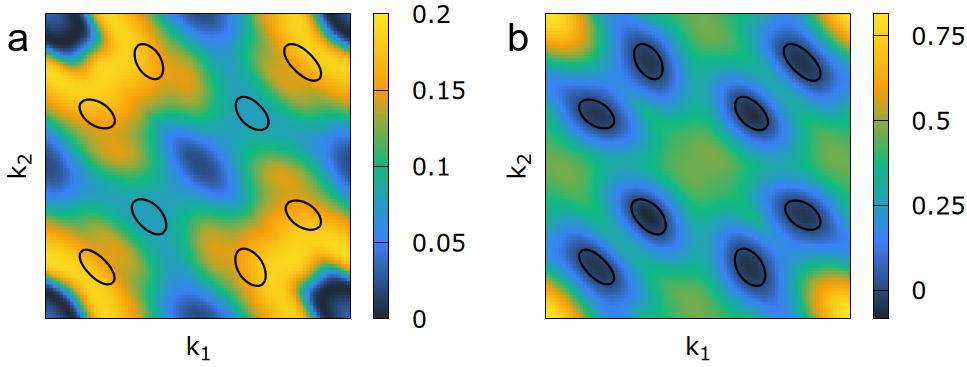}
\caption{Panel a): projection of the Bloch states $p_{n\mathbf{k}}$ over S$_p$ orbitals closer to the disodered potential for the first conduction band as a function of $\mathbf{k}$ point (in reciprocal coordinates) for monolayer MoS$_2$. Panel b): energy (eV) color map of the first conduction band as a function of $\mathbf{k}$. The black lines identify the Fermi level energy surface in both panels. } 
\label{figS1}
\end{figure}

%, where we have two eigenvectors:

%\begin{equation}
%\begin{gathered}
%        \kappa(T_c)^2- \mathrm{Tr} K(T_c) \kappa+\det K(T_c) =0 \\ 
%\Rightarrow \kappa_{max} = \frac12(K_{11}+K_{22})+\frac12\sqrt{(K_{11}+K_{22})^2-4( K_{11}K_{22}-K_{12}K_{21})}
%\end{gathered}
%    \end{equation}

%where the dependence on $T_c$ was suppressed for brevity in the second line. If the second term in the square root is zero one recovers the isotropic solution. Due to the complicated for of the 
\begin{figure}[t]
\includegraphics[width=0.5\columnwidth]{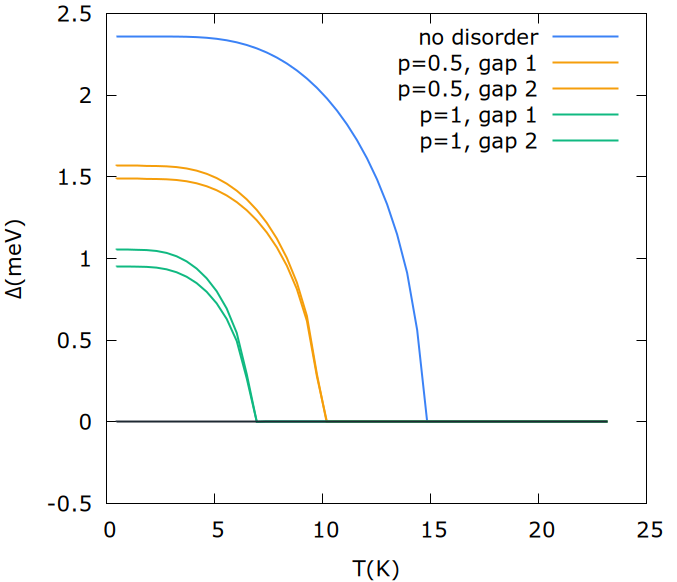}
\caption{Gap reduction and multi-gap character as a function of the disorder parameter $\tilde{x}^\alpha$. } 
\label{figS3}
\end{figure}

\subsection{multi-gap BCS equation in the presence of disorder}

Until now we focused on the linear response regime, where a relationship for $T_c$ could be found also in the multi-valley case. Our equations have been derived employing linear response on the normal state and treating the BCS Hamiltonian as a perturbation. They coincide with the linearized version of the BCS gap equation in the absence of disorder+interaction induced fluctuations. If one wants to extract additional information pertaining to the superconducting gap, one should perform the linear response calculation deep in the superconducting state. We adopt a simpler approach here and leave the calculation of the pair propagator in the superconducting state beyond mean field for a successive work.

We exploit the fact that we already know the BCS kernel in the absence of disorder, $K^0$. Starting from Eq.~\ref{eq:linearBCS} the nonlinear BCS equation is recovered via the substitution $\varepsilon_n^2 \rightarrow E_n^2 =    \varepsilon_n^2+\Delta^2$:

\begin{equation}
    \Delta_\gamma= \sum_{\alpha} \Delta_{\alpha}V^{BCS}_{\gamma\alpha}\sum_{n=1}^{N_{\alpha}} \sum_{m=-\infty}^{\infty} \dfrac{1}{\omega_m^2+E_n^2}
    \label{eq:fullBCS}
\end{equation}

Then, one has to take care of the corrections due to disorder $R_{HF}(T)$ and $R_v(T)$. These are both infrared divergent. A natural way to regularize them and include them in the full BCS equation is by evaluating them employing an infrared cutoff, which is naturally given by the superconducting critical temperature $T_c$. This amounts to consider the following regularized expression for $R^{\alpha}_{HF}(T)$ and $R^{\alpha}_{v}(T)$:

\begin{equation}
\begin{cases}
   \tilde{R}^{\alpha}_{HF,v}(T) = {R}^{\alpha}_{HF,v}(T) \hspace {2cm} T>T_c\\
   \tilde{R}^{\alpha}_{HF,v} = {R}^{\alpha}_{HF,v}(T_c) \hspace {2cm} T\leq T_c
   \end{cases}
\end{equation}

Proceeding in this way we come to the following form for the multi-gap equation in the presence of disorder:

\begin{equation}
    \Delta_\gamma= \sum_{\alpha} \Delta_{\alpha}V^{BCS}_{\gamma\alpha}[\sum_{n=1}^{N_\alpha} \tanh(\dfrac{E_n}{2T})/E_n+\tilde{R}_{HF}^{\alpha}+\tilde{R}^{\alpha}_v]\label{eq:multi-gap_corr}
\end{equation}

This equation is then solved self-consistently in our analysis of the superconducting gap of field-effect doped monolayer and bilayer MoS$_2$. This can be relevant for the superconductivity of MoS$_2$, since the valley at $K$ and $Q$ have different projection over the sulfur states closer to the disordered potential, see Fig.\ref{figS1} a), where we report the sum of the projections of the lowest conduction band eigenstate over the external sulfur states $S_p$, $p_{n\mathbf{k}}=\sqrt{\sum_{i}\braket{w_i}{\psi_{n\mathbf{k}}}^2}$ as a function of the $\mathbf{k}$ point.
In order to show the effect of this correction anisotropy, in Fig.\ref{figS3}
we show the superconducting gap at $T=0$ for a eight band superconductor, where $\theta_D=4$~meV, $\lambda_{ij}=0.11$ if $i\neq j$, $\lambda_{ii}=0$, and the renormalized disorder parameter $\tilde{x}^\alpha=-1/N(0)(\tilde{R}_{HF}^\alpha+\tilde{R}_v^\alpha)$ is only put on six of the eight bands. Interestingly, this mechanism leads to a disorder-induced enhancement of the multi-gap character even in an otherwise single gap superconductor.

\subsection{$\Delta_0$ renormalization in the single valley case}

We discuss Eq.~\ref{eq:multi-gap_corr} in the single valley case:

\begin{equation}
    1 = V^{BCS}N(0)[\int_0^{\theta_D} \dfrac{\tanh(E/2T)}{E}d\varepsilon-x]
\end{equation}

where we replaced the sum over the states by an energy integral and we defined the disorder parameter

\begin{equation}
    x = -1/N(0)(\tilde{R}_{HF}+\tilde{R}_v)
\end{equation}
We are especially interested in finding the relationship $\Delta_0(x)$, where $\Delta_0=\Delta(T=0)$. In this case we do the usual $T\to0$ limit of the BCS gap equation (where $\tanh \to 1$) to find:

\begin{equation}
    1=V^{BCS}N(0)[\ln(\theta_D/\Delta_0)+c-x]\label{eq:0limit}
\end{equation}

where $c$ is a constant. Inverting Eq.~\ref{eq:0limit} one finds the relationship:

\begin{equation}
\Delta_0(x) = \theta_D ~e^c~e^{-1/V^{BCS}N(0)}e^{-x}   \label{eq:delta03} 
\end{equation}

The first observation that we make is that by comparing Eq.~\ref{eq:Tc} and Eq.~\ref{eq:delta03} in our model one finds that the usual $\Delta_0/T_c$ BCS ratio \textit{is unchanged in the presence of disorder}. The second observation is that the correction $x$ \textit{exponentially suppresses the gap} $\Delta_0$. This result has important consequences for our analysis of tunneling spectra, as we discuss in the following section.

\begin{figure}[t]
\includegraphics[width=0.5\columnwidth]{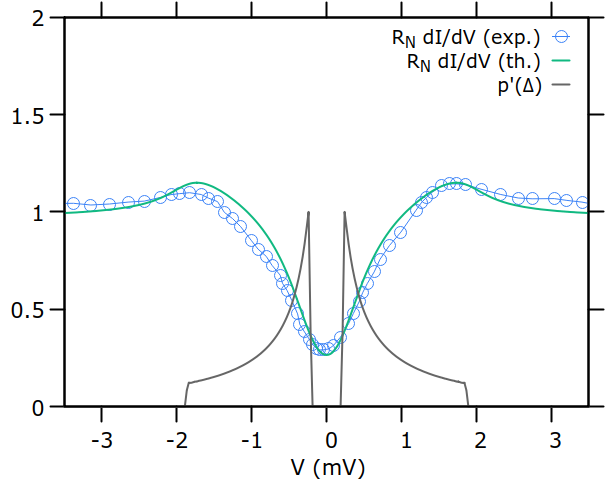}
\caption{Theoretical (green line) vs experimental (blue circles) tunneling conductance with $\sigma=0.01$~meV at T=1.5 K. The grey curve is the gap distribution.} 
\label{figS2}
\end{figure}
\subsection{Theoretical analysis of tunneling conductance and $\Delta$ distribution}

In order to interpret tunneling conductance from Ref.\cite{Costanzo2018} we assume a certain degree of disorder inhomogeneity in the sample. This is modeled in the mathematically simplest way, assuming that the correction $x$ introduced in the previous section is uniformly distributed between two values, $a$ and $b$, according to the relationship:

\begin{equation}
    p(x)=
\begin{cases}
    \dfrac{1}{b-a} \hspace{2cm} a\leq x \leq b\\
     0 \hspace{3cm} \mathrm{elsewhere}
\end{cases}
\end{equation}

This extremely simple choice would correspond to consider a uniformly distributed inverse scattering time, $1/\tau_0$ in the $\{a,b\}$ range, if one neglects the logarithmic dependence. The lowest value $a$ at any doping is naturally chosen to correspond to the $x = -1/N(0)(\tilde{R}_{HF}+\tilde{R}_v)$ value determined as explained in the main text. From this choice for $p(x)$ and resorting to Eq.~\ref{eq:delta03} one can recover a form for the gap probability distribution at $T=0$, $p'(\Delta_0)$, assuming probability conservation:

\begin{equation}
  p'(\Delta_0) =   p(x) |\dfrac{dx}{d\Delta_0}|
\end{equation}

where 
\[ \dfrac{dx}{d\Delta_0} = -\dfrac{1}{\Delta_0}\]

In this case the gap probability distribution becomes:

\begin{equation}
p'(\Delta_0)=
\begin{cases}
    \dfrac{1}{(b-a)\Delta_0} \hspace{2cm} ~e^{-b}\leq x \leq ~e^{-a}\\
     0 \hspace{3cm} \mathrm{elsewhere}
\end{cases}
\end{equation}

This result naturally causes the appearance of a peak in the superconducting gap density of states and can explain the kink feature visible in tunneling spectra of gated transition metal dichalcogenides\cite{Costanzo2018}. 

As already discussed, there is no analytical relationship between the disorder and the gap, $\Delta(x)$, at finite temperature. For this reason, in order to obtain the $\Delta(\varepsilon)$ function we proceed as follows:

\begin{enumerate}
    \item We sample the gap distribution $p'(\Delta)$ by self-consistently solving the $BCS$ equation a large number of times $n$. 
    \item We build the gap function as follows:
    \begin{equation}
        p(\Delta) = 1/n\sum_{i=1}^n \delta(\Delta-\Delta_i)
    \end{equation}
    \item Due to charge inhomogeneity, each Dirac $\delta$ function is substituted with a Gaussian having average $\Delta_i$ and standard deviation $\sigma = 0.3$~meV. The final probability distribution for $\Delta$ is given by:

    \begin{equation}
                p(\Delta) = 1/n\sum_{i=1}^n \mathrm{Gauss}(\Delta,\{\Delta_i,\sigma\})
    \end{equation}
    with $\Delta>=0$, otherwise put to 0. $p(\Delta)$ is renormalized to $1$ if a finite $p$ at $\Delta<0$ is found.
    
    \item The superconducting density of states $N_S(\varepsilon)$ is calculated as:

    \begin{equation}
    N_S(\varepsilon)= N(0)\int d\Delta~p'(\Delta)~\mathrm{Re}\left\{\dfrac{|\varepsilon+i\Gamma|}{\sqrt{(\varepsilon+i\Gamma)^2-\Delta^2}}\right\}\label{eq:sdos2}
\end{equation}
\end{enumerate}

The emergence of a kink feature can be observed in Fig.\ref{figS2}, where the calculated gap probability distribution $p'(\Delta)$ is reported together with the corresponding normalized tunneling conductance, defined in Eq. 10 in the main text. 
The spectra in Fig.2 of the main text are obtained  by solving the BCS gap equations 210 times with a uniformly distributed disorder $x$ between $a=0.2$ and $b=2.3$.

\section{Fermionic and Bosonic corrections to $T_{c}$ in gated TMDs}
\label{app:d}

Our study of the superconducting critical temperature renormalization is based on the assumption that the discussed ``fermion'' fluctuations  related to the enhanced Coulomb repulsion are dominant with respect to boson fluctuation that can reduce the Berezinski–Kosterlitz-Thouless transition temperature T$_{BKS}$\cite{Larkin2008}. A rough comparison between the two is given by the criterion that the "fermionic" critical conductance $g_c^F = (\dfrac{1}{2\pi} \ln\dfrac{1}{T_{c0} \tau_0})^2$ is larger than the bosonic one, $g_c^B=2/\pi$\cite{Larkin2008}. This leads to the condition

\begin{equation}
    \ln\dfrac{1}{T_{c0}\tau_0}>5
\end{equation}

for the fermionic mechanism to be dominant and $    \ln\dfrac{1}{T_{c0}\tau_0}<4$ for the bosonic one to prevail. In the cases studied here one obtains values $\ln\dfrac{1}{T_{c0}\tau_0} \approx 2-3$. However one must take into account the multi-valley character of MoS$_2$. Due to the valley degeneracy and in an isotropic assumption, the critical fermionic conductance becomes

\begin{equation} g_c^F = (\dfrac{4}{\pi} \ln\dfrac{1}{T_{c0} \tau_0})^2\end{equation}

and thus the condition for the fermion $T_{c}$ suppression mechanism to be dominant becomes

\begin{equation}
    \ln\dfrac{1}{T_{c0}\tau_0}>5/8
\end{equation}

which is always the case in the considered scenarios. Incidentally, the relatively low values of the logarithm in this system means that the Hartree-Fock correction $R_{HF}$ cannot be neglected.

\section{Disorder correction to $T_c$ in a 2D homogeneous electron gas}
\label{app:f}

\begin{figure*}[b]
\includegraphics[width=0.9\columnwidth]{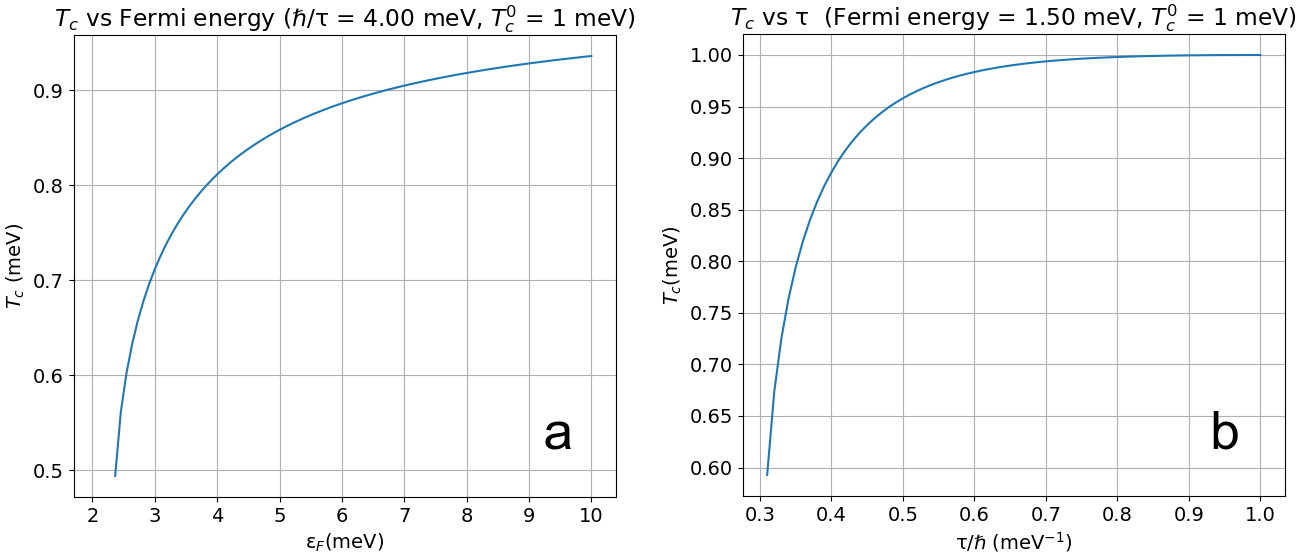}
\caption{Suppression of critical temperature $T_{c}$ as a function of increasing Fermi energy (left panel). Suppression of critical temperature $T_{c}$ as a function of increasing scattering time (right panel).} 
\label{figS9}
\end{figure*}

\section{The monolayer case}
\label{app:mono}

Monolayer samples show very low superconducting critical temperatures (T$_c\approx 2$ K or no superconductivity \cite{Yecomm,Costanzo2016,Fu2017}) with respect to their few-layer counterparts\cite{Costanzo2016}. We showed in Sec.\ref{app:a} that this is can be partially attributed to a stronger suppression of conductivity.

\begin{table}[htp]
    \centering
    \begin{tabular}{|c|c|c|c|c|c|}
         \hline
  system  & n$_e$/cell  & R$^0_{\square}(\Omega)$ &  $\varepsilon_F^{avg}$ & $\tau_0$ (fs)  &  $R_{HF+v}(T_c)/N(0)$  \\
         \hline 
                  \hline
           monolayer & 0.1  & 85.02 & 0.05 & 109.1  & -0.149  \\
           \hline
          bilayer & 0.1 &  52.49  & 0.05 & 169.85 & -0.064 \\
\hline
          bilayer & 0.125 &  82.49 & 0.06 & 108.08 & -0.135 \\
          \hline
           bilayer & 0.175  & 139.71 & 0.08 & 50.56 & -0.733 \\
         \hline
    \end{tabular}
    \caption{Model-derived parameters entering the self consistent formula for $T_{c}$ in the presence of disorder.} \label{tab1}
    \end{table}

To be more quantitative, we also applied our model to the single-side gated monolayer system, employing the value $T_{c0}=15$~K at $n_e=0.1 e^-$. The obtained parameters for the monolayer are reported in Tab.\ref{tab1}, together with the studied cases in single-side gated bilayer. The model predicts $\approx$ 14\% renormalization for $T_{c}$ in monolayer at $n_e=0.1/$~cell. This indicates that the effect of gate-induced disorder is stronger than in the bilayer case at similar doping, but it is not sufficient to explain the almost absent superconductivity in monolayer\cite{Costanzo2016}. This fact is only apparently surprising: even if the induced charge is slightly more spread off in the multilayer than in the monolayer (see right panel of Fig.\ref{figS6}), a large (>90\%) fraction of it is always found in the first layer. A natural explanation for the large $T_c$ suppression in monolayer\cite{Costanzo2016,Fu2017,Ding2022,PhysRevB.1.1078} is the concomitant effect of gate- and substrate-induced disorder. Since gate-induced superconductivity is localized in the monolayer directly in contact with the gate independently from the number of dichalcogenide layers, all the layer beyond the first play the role of a buffer, preventing the substrate from disrupting superconductivity through additional scattering mechanisms. The bilayer in this case represents an intermediate situation where only a small part of the induced charge is directly in contact with the substrate. In the end, we expect monolayer to be characterized by an effectively higher resistance consistently with experiments\cite{Fu2017}, while superconductivity would be protected by the extra layers in the bilayer (partially) and in thicker samples.

\section{Comparison with previously published calculations}
\label{app:g}

\begin{figure*}[b]
\includegraphics[width=0.45\columnwidth]{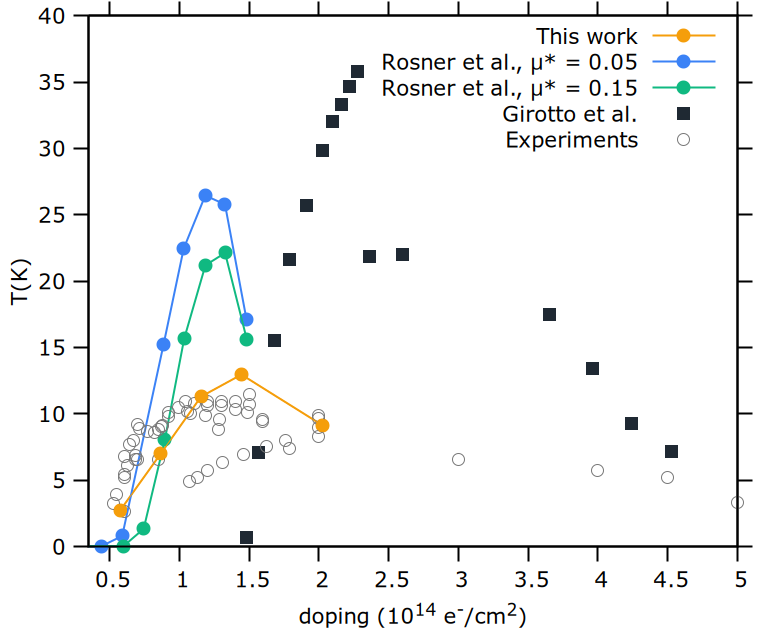}
\caption{Predicted superconducting critical temperature from the present work is here compared to previously published calculations from Refs.\cite{PhysRevB.90.245105, GirottoErhardt2025} and experimental data taken from Refs.\cite{doi:10.1021/acs.nanolett.8b01390,Costanzo2018,doi:10.1126/science.1228006,Saito2016,doi:10.1126/science.aab2277,AliElYumin2019}.} 
\label{figS8}
\end{figure*}

In this section we report a comparison between previous calculations and the present results for the $T_c$ vs doping curve. The results are depicted in Fig.\ref{figS8}. The  dome reported in Ref.\cite{PhysRevB.90.245105} is a spurious nonphysical consequence of the Allen-Dynes formula application at high values of $\lambda$. The dome in Ref.\cite{GirottoErhardt2025} is interpreted as due to charge-density wave formation in MoS$_2$, contrary to the conclusions reached in the present work and in Ref.\cite{Marini_2023}.

\section{Computational details}
\label{app:e}

Transport simulations in the presence of disorder are performed within the model described in Appendix~\ref{app:a} and \ref{app:b}, starting from a first principles calculation and generating a tight binding model from the Wannierization. We employ a $6\times6$ supercell to simulate conductivity in the disordered system, containing 108 atoms for the monolayer and 216 for the bilayer dichalcogenides. Lattice parameters are fixed to the experimental ones as in Ref.\cite{Marini_2023}, internal coordinates are relaxed in the presence of the homogeneously charged plate(s)\cite{PhysRevB.96.075448}. Doping induced lattice expansion is not included in the calculation. Positively charged impurities are included on top of gating according to the model presented in Appendix~\ref{app:b}, assuming a vertical distance between the charged DEME$^+$ molecular impurities and the chalcogen atoms of 5 \AA~ for MoS$_2$ and $5.1$ \AA~, as discussed in the main text. The relative permittivity of the environment $\kappa$ was set to 2.5. With this value of $\kappa$, the Thomas-Fermi screening parameter $q_s$  was calculated to be $\approx$ 8.35 \AA$^{-1}$ at a doping of 1.14 e$^-$/cm$^{-2}$ in single-side gated bilayer MoS$_2$.~ The numerical solution of Eq.~\ref{eq:poisson2} was performed expressing the Hankel transform of the second order differential equation Eq.~\ref{eq:poisson2} discretely and using the finite difference formula for the second derivative to express it as a tridiagonal system, which was then solved employing Thomas algorithm at every $q$. A certain number positively charged impurities were included in the periodic supercell calculation, so to match the target impurity concentration $n_{imp}$. For the Hankel transform we used a maximum $q_{max}=10$  \AA$^{-1}$ and $z_{max,min}=\pm10$ \AA. The model was further simplified by assuming that the$\omega_{i,\mathbf{R}}$ are spheres of radius $r_w=2$\AA ~centered on the atom to which the Wannier function belongs, mimicking the position of the free carriers. Thus we substitute to the integral in Eq.\ref{eq:int} the average value:

\begin{equation}
    U_i(\mathbf{R})\approx\dfrac{U_i(\mathbf{R_{\parallel}},z=z^i_{max})+U_i(\mathbf{R_{\parallel}},z=z^i_{min})}{2}
\end{equation}
$z^i_{max}$ and $z^i_{min}$ being the positions closest to and farthest from the gate within the considered sphere and $\mathbf{R}_{\parallel}$ the position in the in-plane direction. To avoid the spurious overscreening of disorder on $Mo$-centered states due to Thomas-Fermi model, the Mo-centered Wannier functions were shifted upward to align with the S-centered orbitals, and the additional screening was incorporated by reducing the disorder potential to two-thirds of its original value. To accelerate the calculation, we only include contributions up to $r_{max}=16$ \AA~in the radial direction, after which the induced potential is negligible. 

\begin{figure*}[t]
\includegraphics[width=0.45\columnwidth]{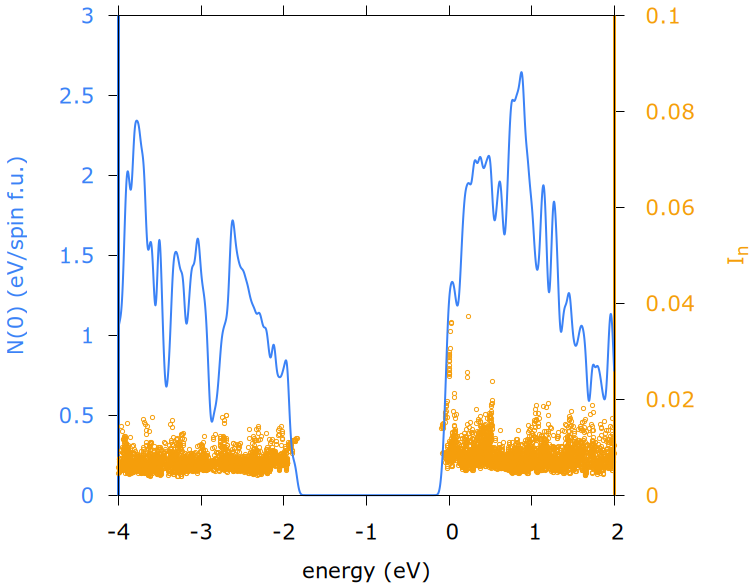}
\includegraphics[width=0.45\columnwidth]{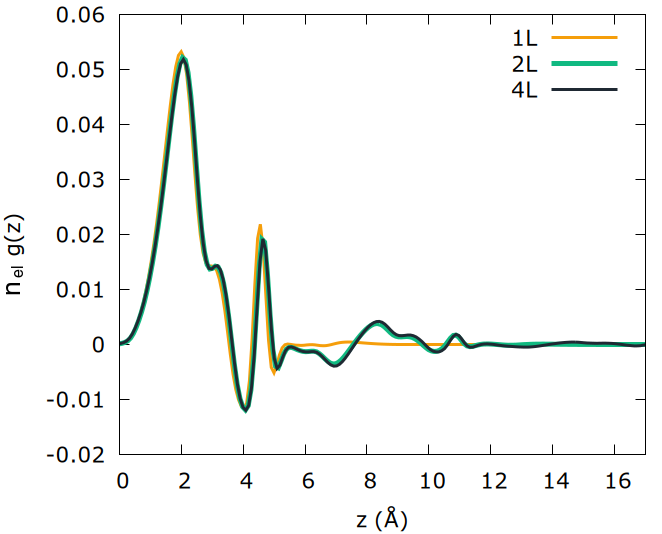}
\caption{Left panel: density of states and inverse participation ratio $I_n$ for a generic configuration in monolayer MoS$_2$ at $n_{imp}=0.166$/cell. Right panel: Planar average of the induced charge $n_{el}~g(z)$ for single-side gated monolayer, bilayer and four layer system. } 
\label{figS6}
\end{figure*}

~To calculate the frequency dependent $\sigma_{2D}(\omega)$, we average on disorder assuming random in-plane positions for the impurity. 10 to 30 disorder configurations were employed to converge $\sigma_{2D}(\omega)$ in each system. Spin-orbit coupling is neglected in the calculation of sheet resistivity. Doping induced lattice expansion is not included in our calculations. For the calculation of conductivity, a broadening of $\delta=0.025$~eV and an electronic temperature of $T_{el}=~0.01$~eV were adopted in the conductivity calculation to accelerate the convergence. The Brillouin zone of the supercell was sampled employing a $6\times6$ Monkhorst-Pack grid. For the calculation of resistivity in the absence of impurities, $72\times72$ Monkhorst-Pack grids were employed to converge the calculation. The disorder-induced renormalization of the single particle density of state $N(0)$ were and not included in the evaluation of $\tau_0$ from R$_\square$. Consistently, the same effect is not included in the renormalization of the electron-phonon coupling $\lambda$. 

To evaluate the localization of electronic states in disordered system, in the left panel of Fig.\ref{figS6} we report the density of states (blue line) and the inverse participation ratio for the state $n$, $I_n$ (orange circles), defined as

\begin{equation}
    {I}_n = \sum_{\mathbf{R},i} |\braket{w_{\mathbf{R},i}}{\psi_n}|^4
\end{equation}

where the sum is extended to all orbitals in the cell. $p_r$ is a measure of the electronic state localization. We observe the formation of a localization tail in the conduction band, however the participation ratio is still small. We find that the tail is absent in the absence of impurities and present in the disordered configurations. The self consistent solution for $T_c$ renormalization was performed graphically. The multi-gap BCS equation was solved self-consistently fixing a Debye temperature $\theta_D=4$~meV, which gives results in good agreement with the first principles calculations of Ref.\cite{Marini_2023}. The tunneling spectra were obtained from the procedure described in Sec.\ref{app:c} G. , with $\sigma = 0.3$ meV and $\Gamma = 0.01$~meV. 

Finally, in the right panel of Fig.\ref{figS6} we show the planar average of the induced charge $n_{el}~g(z)$ as a function of the number of layers. Here, it is apparent how the charge confinement only changes significantly from the monolayer to the bilayer case, while in the case of an higher number of layers  ($>3$) the profile it is almost identical to the bilayer. This is consistent with what we found in Ref.\cite{Marini_2023} for the gated electronic band structure, which was found to be almost identical between the bilayer and the four-layer system.

\bibliography{bibliography}

\end{document}